# Beyond Value Elicitation: Towards Moral Profiles in Early Requirements Engineering via Role-Playing Games and Anthropologist LLMs

Gianluca De Ninno. Computer Science Area, Gran Sasso Science Institute, L'Aquila, Italy and Department of Computer Science, University of Pisa, Pisa, Italy.
gianluca.deninno@gssi.it

Paola Inverardi. Computer Science Area, Gran Sasso Science Institute, L'Aquila, Italy.
paola.inverardi@gssi.it

Francesca Belotti. Department of Humanities, University of L'Aquila, L'Aquila, Italy.
francesca.belotti@univaq.it

## Abstract

This study presents a proof of concept for eliciting and representing the moral profiles of digital system users in Requirements Engineering (RE) by combining immersive role-playing games (RPGs) with large language model (LLM) analysis. While existing approaches rely on predefined value taxonomies and explicit articulation, values are often tacit, context-dependent, and difficult to express directly. To address these limitations, we propose moving from the elicitation of discrete moral values to the narrative reconstruction and representation of users' moral profiles. Grounded in phenomenological and narrative anthropology, the approach focuses on capturing users' moral orientations as they emerge through situated decision-making. RPG sessions generate context-rich narrative data, which are then analyzed by a specialized LLM (GPT-A) to produce individual anthropological moral profiles (IAMPs). A validation process based on cross-comparison between model outputs and participants' responses in unseen moral scenarios assesses the adequacy of the generated representations. Results indicate that RPG environments effectively support the generation of rich, context-dependent data for eliciting tacit values, and that an anthropologically grounded LLM can transform such data into coherent narrative representations of users' moral profiles. These representations enable the contextual interpretation of users' preferences and values within the given domain, with improved performance when interpretive framing captures relationships between actions, underlying motivations, and individual domain expertise. From an RE perspective, this approach enables the analysis of user preferences and trade-offs while preserving their situated and dynamic nature, providing a foundation for integrating human moral values into the early stages of requirements engineering.

## INTRODUCTION

The increasing reliance on software systems, particularly autonomous systems, has introduced significant ethical, social, and political challenges, including issues of fairness, accountability, and societal bias [1]. As a result, there is a growing call from researchers to systematically include human values into software systems [2].

In Software Engineering (SE), the process of identifying, formalizing, implementing, and verifying values within software systems is known as operationalizing human values [3]. In our study, we focus on the elicitation phase because, at this stage, the software is still malleable, making it essential to effectively developing value-oriented software [4, 5].

Various methods and value models have been proposed to integrate human values throughout the Software Development Life Cycle (SDLC) [6]. These approaches assume that users' moral values can be categorized into a finite set of universal categories based on established value models. Although this perspective is useful for large-scale analyses or comparative studies, it gives rise to three main practical challenges: first, values are inherently subjective and often tacit, making them difficult to interpret outside their specific social and cultural contexts; second, values change over time; and third, there is still a lack of practical methods and value-oriented guidelines to effectively elicit human values [6].

Traditionally, RE focuses on functional and non-functional requirements while paying comparatively limited attention to human-related ones, including stakeholders' values and emotions [4]. In recent years, researchers have increasingly explored the integration of human-centered requirements into RE, addressing aspects such as emotional responses [7, 8], gender differences [9, 10], power dynamics [11], personality traits [12, 13], age-related factors [14], and both mental and physical conditions [15]. However, incorporating such values into software design remains challenging. Key difficulties include the absence of clear guidelines, the complexity of translating abstract value concepts into concrete development tasks, the lack of measures to assess how values are reflected in software systems, and the limited availability of approaches to trace these values from initial elicitation through to the final delivery of the product [16, 17].

This work addresses three key challenges: (1) overcoming limitations of taxonomy-based approaches (2) eliciting tacit and context-dependent individual moral values, orientations and preferences, (3) provide narrative description of the elicited values in terms of moral profile. In doing so, it also facilitates communication between stakeholders and elicitors. In our study, we distinguished moral values between hard ethics values and soft ethics ones, based on Floridi's characterization of digital ethics [18]. We focus on the latter because hard ethics is directly shaped by laws and regulations, which define moral duties, rights, and responsibilities, determining what is legally permissible, whereas soft ethics operates beyond legal compliance, guiding human behavior through self-regulation rather than through legislation.

On the other hand, in this paper we present an approach to moral values, preferences and orientations grounded in narrative and phenomenological anthropology, assuming that they can't be redacted to universal categories, are strictly linked with the context in which they emerge and are shaped by lived experience, social interactions, and can be investigated through the narratives individuals produce about their moral lives, being continuously reinterpreted through personal and collective meaning-making [19-22]. Accordingly, the analysis does not focus on explicitly categorizing users' moral values; rather, moral profiles are inferred by analysing choices made in context, their justifications, and the narrative framing of those decisions.

Our approach combines ethnographic role-playing simulation to elicit moral decision-making in context and LLMs to generate representation of participants' moral profiles. Borrowed from game study, it involves immersive role-playing games (RPGs) as an ethnographic tool where the researcher guides the gameplay experience, while participants act themselves in a realistic setting. Immersive RPG was chosen due to its capability to provide a controlled, offering an alternative to traditional elicitation techniques such as interviews and surveys, as it is structured around making observable moral orientations as they emerge and evolve during gameplay [23], and due to its inherently discursive and storytelling-based nature fosters continuous interaction among players, making the process of moral reasoning explicit through dialogue [24]. LLMs were selected for their ability to interpret complex qualitative and narrative data, and to support the scalability of the proposed protocol. Recent studies show that LLMs can effectively analyze textual data, including narrative materials, producing results comparable to human qualitative analysis while also offering complementary insights [25, 26]. At the same time, their integration into qualitative workflows has been shown to significantly reduce the effort required for tasks such as thematic analysis, enabling qualitative approaches to be applied at scale [27]. However, LLMs are not intended to replace ethnographic analysis, but to complement it within a human-in-the-loop protocol that preserves interpretive depth while extending analytical capacity.

The collected data consist of a combination of players' self-disclosure data, self-ethnographic multimedia annotations, and researchers' ethnographic fieldnotes, all gathered during the game sessions and follow-up sessions. The gathered data were systematized into coding tables (*narrative tables*) that relate scenarios characteristics to the decisions made, as well as to the in-game and off-game factors that influenced the final decision. These narrative tables, together with contextual information and an anthropological analytical framework, were provided to ChatGPT-4o (hereafter referred to as *GPT-A*). GPT-A was asked to generate an individual anthropological moral profile (*IAMP*) for each participant that constitutes the model's moral profile narrative representation—an implicit representation of elicited moral values rather than a list of explicit ones.
To validate this approach, three main research questions are raised:

- **RQ1:** Are the narrative data generated in RPG sessions sufficiently rich and informative for the implicit elicitation of the user's moral profile?
- **RQ2:** To what extent can Anthropologist LLMs support the generation of narrative representation of elicited moral values?
- **RQ3:** Can the combination of RPGs and a customized LLM efficiently improve the comprehension of the users' needs, preferences and values?

To establish a reliable benchmark, we implemented a two-level cross-validation system. The first level is based on a comparison between the model's responses and those of the participants in a subsequent questionnaire built on unforeseen scenarios involving moral dilemmas in the same ethical domain of the RPG, aimed at evaluating the model's ability to capture users' moral orientations based on the generated individual moral profiles.The second level consists of comparing different GPT configurations to evaluate differences in their accuracy in predicting users' responses in the aforementioned unforeseen scenarios, i.e., through an interpretative framing test. The full methodology—game design, data collection and analysis, GPT configuration, and validation—is outlined in the following sections.

## BACKGROUND

In this section, we review how values are conceptualized in Computer Science (CS) and the most commonly used methods and tools for values elicitation in RE. We then introduce the anthropological concepts and methods that inform our approach, and examine prior work on the use of games in RE. Finally, we analyse the specific characteristics of RPGs that justify their use as a value elicitation tool and the capability of LLM to analyse qualitative data based on previous works.

*Human Values and Elicitation Techniques in RE*

Human values are broadly defined as “what is important for an individual or a society” [28] and as principles that guide human action in everyday life [29]. Value models aim to identify universal values shared across human individuals. This classification process is useful for defining reference categories that help classify stakeholders’ specific values within particular contexts and domains. One well-established value model is that proposed by Schwartz et al. [30], which defines ten core values arranged in a circular structure. In recent years, this approach has been refined, and the number of values has increased through the integration of modern theories of motivation, suggesting that contextual motivations and emotions are fundamental drivers of human decision-making [31]. Indeed, the interplay between these factors provides insight into why individuals act in specific ways, such as approving or disapproving of certain actions and choosing to engage or disengage in specific activities [32].

A pioneering attempt to incorporate human values into technological systems is Value-Sensitive Design (VSD), proposed by Friedman et al., which “accounts for human values in a principled and systematic manner throughout the design process” [33]. VSD adopts an interactional stance, emphasizing the relationship between technology and human values. However, VSD faces several practical challenges in real-world applications [34]: it struggles to capture context-dependent (*situated*) values, provides limited support for resolving value conflicts, and lacks mechanisms to account for the evolution of values over time. Many other attempts have been made to embed human values into SDLC. For instance, Value-Based Requirement Engineering (VBRE) provides a step-by-step methodology to guide developers in identifying stakeholders’ values from different requirement engineering artifacts, documents and interviews [35]. This method relies on a paper-based taxonomy of values, employing emotion-focused surveys and interviews to capture stakeholders’ emotional traits [36]. Koch et al. proposed a method for inferring end-user values by analyzing their task preferences within a specific work environment [37]. Similarly, Lopez et al. introduce a workshop-based methodology that utilizes tools such as value cards to facilitate discussions among developers and stakeholders [38]. By engaging in structured conversations, developers can gain a deeper appreciation of the ethical and practical considerations related to software security. Perera et al. [39] proposed the use of Persona for stakeholders’ values identification. In the study, human values are linked to system features across different levels of the SDLC.

These approaches offer several advantages, facilitating large-scale analysis, consistency, and comparative evaluation throughout the development process [40]. However, several important limitations must be acknowledged. RE is often affected by communication breakdowns and asymmetries in knowledge exchange between stakeholders and elicitors, which can leave relevant information only partially articulated, unexpressed, or entirely unrecognized [41, 42]. Moreover, encoding values into a fixed ontology or a single metric can oversimplify complex ethical considerations. Selbst et al. [43], studying fairness in machine learning algorithms, call this the "Formalism Trap": when designers treat a formal definition as exhaustive, they may assume a system is "fair" or "secure" based solely on pre-defined metrics, whereas real-world stakeholders may contest these claims. Albeit offering clarity and consistency, quantitative approaches might raise ethical concerns related to reductionism creating a false sense of objectivity around both subjective or pluralistic values.

These limitations highlight the need for alternative elicitation approaches that do not strip away values from the context in which they emerge and do not oversimplify complex ethical concerns into predefined taxonomies.

*From Moral Values Taxonomies to Moral Narrative Profile: an Anthropological Shift*

Between the 1980s and 1990s, anthropology underwent a significant shift in its approaches, with phenomenological anthropology and narrative anthropology gaining prominence. The former emphasizes embodiment and lived experience, whereas the latter focuses on storytelling as a fundamental human practice.

The phenomenological perspective has its roots in the philosophy of Merleau-Ponty, who argued that lived experience is shaped by culture and is inherently interpretative [44]. Kleinman (1995) argued that both individual and collective moral life take shape within specific cultural, relational, and historical contexts, where particular goods, responsibilities, and vulnerabilities are at stake [19]. Kleinman introduced the concept of the *local moral world*, where "moral" refers not only to the values themselves, but also to the meanings they acquire in a collective and individual perspective. In this view, moral values are not stable entities to be elicited as such; rather, they often remain unarticulated until individuals face situations of uncertainty, conflict, or tension [45]. These moments of disruption do not merely uncover pre-existing hierarchies of values; rather, they highlight how moral orientations are dynamically constructed in response to contextual challenges. As also noted in the RE community [43], these observations suggest that predefined categories risk reducing reality to static representations.
Narrative anthropology emphasizes that moral life become meaningful through storytelling, as individuals interpret experiences, justify their actions, and evaluate possible alternatives [20, 22]. Since narratives are co-constructed through interaction, they reveal the inherently dialogical and relational nature of moral values [21]. This is particularly relevant in contexts where values are not directly articulated: storytelling provides a medium through which tacit knowledge can be expressed, negotiated, and shared. Furthermore, because individuals are situated within social, cultural, and relational contexts, the expression of moral life is not entirely unconstrained but shaped by shared systems of meaning [46, 47].
Methodologically, narrative anthropologists often use in-depth interviewing and participant observation: the former serve to aggregate personal experiences, stimulate reflective thinking, and generate moments of learning [48], whereas the latter involves "extended immersion in a culture and participation in its day-to-day activities" [49]. Participants' stories are then analyzed both in terms of content and form: researchers examine what is said (themes, cultural symbols, moral messages) and how it is said (structure, style, audience interaction).
In this research we interpret values as relational and narrative patterns expressed through behavior. In this sense, behavior is a situated enactment of moral life that emerges from lived experience, cultural meanings and social negotiation. For this reason, the focus of the proposed elicitation protocol is the process through which individuals interpret situations, negotiate meanings, and orient their actions in response to what is at stake. Moreover, adopting such an approach might address communication challenges by focusing on inferring co-created values based on moral profile expressed and constructed through narratives, interactions, and justifications, rather than assuming they can be directly articulated.

*Serious Games and Role-Playing Games*

The use of games in knowledge production has expanded significantly in recent decades [50], leading to the emergence of "gamification science"—an interdisciplinary field that investigates the theories, methodologies, and empirical effects of incorporating game mechanics into various contexts. Unlike basic gamification, which simply applies game-like elements to designed tasks, gamification science is a structured research approach that studies the cognitive, emotional, and social dimensions of game-based interventions [51]. A major development within gamification science is the use of "serious games". This term refers to those games that have "an explicit and carefully thought-out educational purpose that are not intended to be played primarily for amusement" [52]. They can be analog or digital and are designed with educational [53], training [54] or data collection purposes [55].
Among serious games employed in gamification science, there are RPGs both in the analogical (TTRRPGs) and in digital form (MMORPGs) [56]. In RPGs each player acts as a fictional or realistic character in an imagined world. RPGs have rules that define how, when, and to what extent each player can influence the imagined space and interact with it. A Game Master (GM) facilitates the experience by shaping the narrative, describing the environment, portraying non-player characters (NPCs), and managing the consequences of players' choices. During the game, players may encounter NPCs that are integrated into the game world, each with their own personalities, motivations, and roles within the story. NPCs can range from allies and quest-givers to antagonists, and they serve to enrich the narrative, provide information, create challenges, and make the world feel more dynamic and believable.
We conceptualize RPGs as setting based on storytelling that shapes and redefines individual and collective narratives [24]: as players justify their choices and negotiate their consequences, values, needs and beliefs are enacted and made publicly accountable through interaction, continuously reshaping moral thresholds and bringing implicit norms to the surface [23, 24, 57]. Indeed, a core feature of RPGs is player/character identification, which fosters emotional and cognitive engagement and enhances immersion through storytelling, character development, and collaborative problem-solving [23, 57, 58]. Unlike other games, RPGs are characterized by unpredictability [59], reflecting real-world complexity and enabling high levels of engagement [60, 61].
Due to its characteristics, RPGs are particularly suitable for addressing the limitations of traditional values elicitation methods in RE, as they enable the observation of situated decision-making processes through which it becomes possible to identify the contextual elements that shape users' moral decisions and, by taking these into account, reconstruct their moral profiles and infer their moral orientations and the values enacted across different situations and contingencies.
Therefore, as highlighted by Dalpiaz and Cooper [62], the effectiveness of game-based methods in RE depends on careful design and evaluation. Particular attention should be paid to game design to avoid superficial participation, as well as to the upfront estimation of development, training, and integration costs. Nevertheless, the use of gamified methods or serious games is recognized in RE for its ability to yield positive results in several aspects [63]. One of the biggest advantages is that games make the elicitation process more interactive, encouraging individuals to actively engage rather than passively answering questions [64, 65]. Additionally, games stimulate creativity and communication, leading to more diverse and insightful requirements [66]. They foster collaboration by creating a structured environment where individuals can discuss and resolve conflicts, and they provide real-time feedback, allowing for immediate refinement of ideas [67, 68].

*LLMs for elicitation tasks in RE*

Recent advances in generative AI, and in LLMs, have introduced new possibilities for supporting RE activities. Current research shows that LLMs can assist across multiple RE activities, including requirements elicitation, specification, validation, and traceability, by supporting tasks such as requirements generation, quality assessment, and classification [69]. At the same time, their integration in RE remains at an early stage, with most studies conducted in controlled environments and limited industrial adoption, highlighting both their potential and the need for more systematic approaches to their use in practice [70].

Among RE activities, elicitation has emerged as one of the primary areas where LLMs can provide support [71]. Elicitation traditionally relies on interaction-intensive techniques to uncover stakeholders' needs. LLMs introduce new forms of support in this phase by assisting in the generation of elicitation materials (e.g., interview questions) [72], analyzing the unstructured materials from interviews and refining users' story for requirements generation [73]. A widely explored approach technique for employing LLMs in elicitation is role-play or prompt-based workflows, where LLMs simulate stakeholders to generate needs or plausible scenarios [17, 74, 75]. In these approaches, role-playing is primarily used as a prompting or interaction technique. In contrast, our research adopts a role-playing game–based elicitation setting, in which human participants—rather than LLMs—engage in a structured, game-based role-play activity.

Traditionally unstructured data can be coded through different qualitative methods, relying on coding, thematic analysis, and interpretive reasoning, which are labor-intensive and require significant human effort to identify patterns, themes, and narratives in textual data [76, 77]. Recent work shows that LLMs can support these processes by automatically identifying recurring patterns, assisting in the construction of themes, and analyzing narrative data often achieving results comparable to human analysts while significantly reducing time and effort [25, 26, 78]. Rather than replacing human interpretation, current approaches increasingly frame LLMs as collaborative analytical tools, supporting iterative refinement within human-in-the-loop workflows [79]. However, several challenges remain, including issues of interpretability, potential bias, and the risk of generating plausible but inaccurate interpretations [79]. For this reason, recent work emphasizes the importance of transparent prompting strategies, explicit analytical frameworks, and validation procedures to ensure the reliability and credibility of LLM-assisted qualitative analysis [80].

Building on this line of research, our work leverages LLMs to support structured qualitative interpretation beyond coding automation, reconceptualizing their role as an interpretive partner operating within a theoretically grounded and human-supervised framework [81]. By combining LLM-based analysis with an anthropological perspective, we move beyond coding acceleration toward a richer, context-sensitive interpretation of users' moral profiles, as they emerge from the data through explicit analytical lenses.

## METHODOLOGY AND STUDY DESIGN

Starting from the redefinition of values as part of individual local moral world through the narrative and phenomenological anthropology lens, we design a protocol for moral profile elicitation and representation divided into four main steps:

1. constructing the environment for moral profiles elicitation. It includes the definition of PRG's domain; scenarios generation following specific design rules; the definition of the moral dilemma of each scenario;
2. data collection through gameplay, participant observation and follow-up sessions (i.e., focus-group);
3. Individual narrative moral profile generation through an LLM-assisted process. This phase relies on the anthropological analysis of the collected data, supported by a specialized GPT model informed by phenomenological and narrative anthropology, a glossary of relevant anthropological concepts, and materials describing the game design rules and overall structure. The output of this process is the individual anthropological moral profile (IAMP) for each participant;
4. the validation of the IAMP. To assess whether the protocol captures meaningful representations of users' moral orientations, we generate new scenarios within the same domain and ask participants to make decisions in these contexts. The responses of each player are then compared with those predicted by the model based on the IAMP. The degree of predictive accuracy is used as an indicator of the quality of the elicitation process.

When misalignments occur, the profiles should be iteratively refined through additional feedback and data collection. However, this study focuses only on the initial elicitation and validation cycle, in order to assess whether the combination of RPGs elicitation and LLMs analysis can capture meaningful moral profile representations. The proposed protocol received ethical approval from our institutional MIRB (reference number 186709).
The whole protocol is presented in Fig 1, the dashed line represents an iterative refinement step not applied in this study.

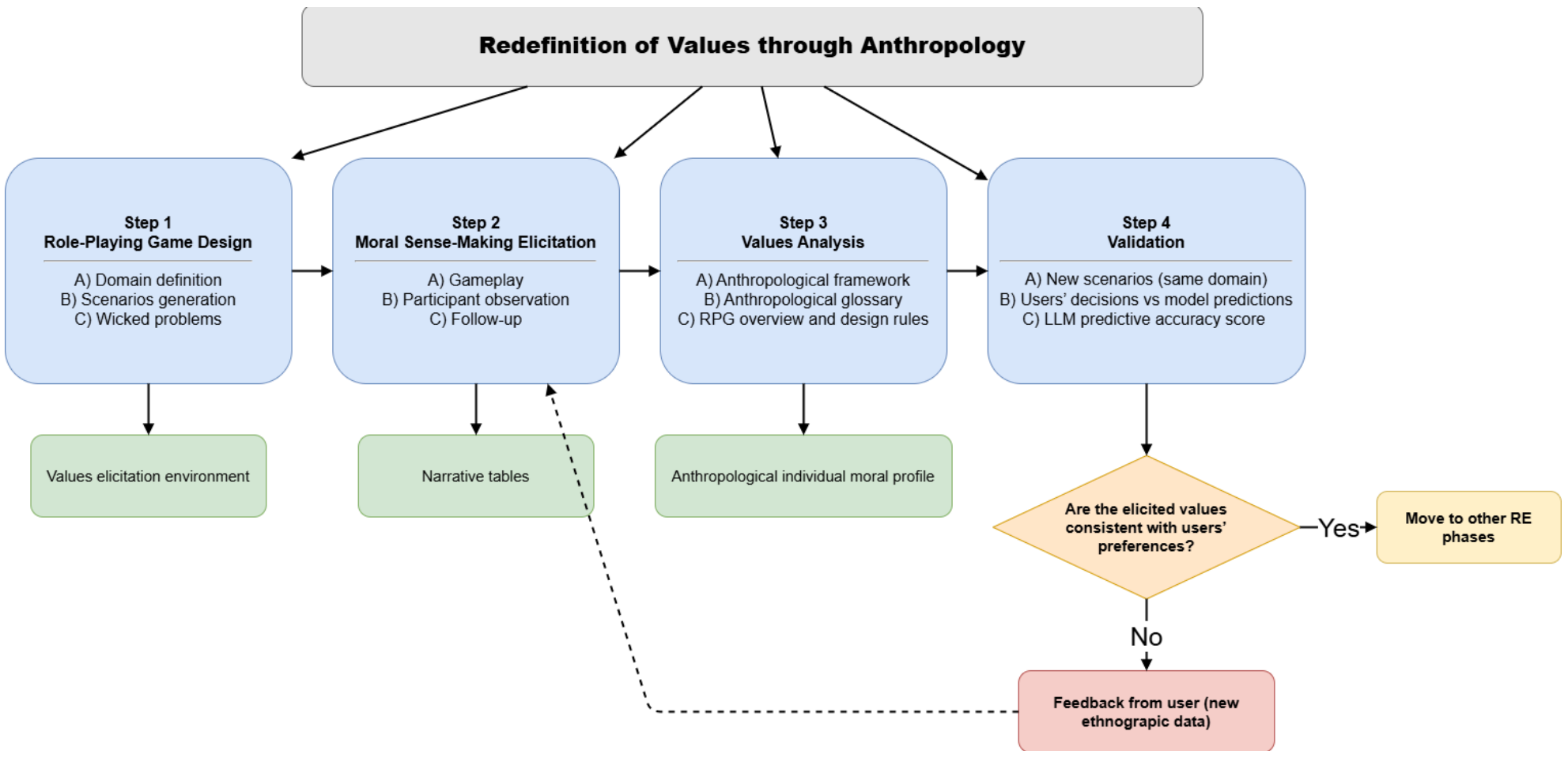

Fig. 1 Overview of the proposed protocol.

The ethical domain chosen for this proof-of-concept is digital privacy. This domain was selected due to its increasing relevance in the era of automation [82, 83], as well as its widespread salience among users of digital technologies [84], thereby minimizing the risk of participants misunderstanding the issue or requiring specific technical expertise to engage with it.
Accordingly, the case study focuses on analysing how privacy related moral values and orientations emerge and can be elicited through the proposed protocol. While privacy serves as the target domain in this study, the protocol is designed to be domain-agnostic and can be readily applied to other ethical contexts (e.g., human dignity in healthcare, sustainability, ethical business practices) by adapting the scenarios while preserving the underlying mechanics.
Selected examples of the game scenarios, evaluation questionnaire, and materials used to personalize GPT-A are presented in the following sections; the complete set of materials is provided in the Supplementary Materials.

*Phase 1: Game Design and Gameplay*

In the RPG designed for the study each player acts themself, while the researcher acts as the Game Master (GM), facilitating the progression of the story. The game consists of eight structured scenarios, followed by a final scenario that resolves the narrative. Because role-playing games are inherently open-ended and can otherwise lead to highly variable and extended gameplay sessions, the number of scenarios was deliberately limited to constrain the duration of the activity to one to two hours. This design choice helped balance meaningful engagement with manageable cognitive load and supported the reproducibility and control of the experimental protocol. An example of a scenario is presented in Fig. 2.

| Scenario #5 | |
|---|---|
| Description | Upon exiting the building, the players encounter a concierge. He requests assistance in setting up a Google account, creating a secure password, using email securely, and navigating the internet without falling victim to fraudulent advertisements. |
| Wicked problem | **Should they allocate time and expertise to assist him or ignore the request?** |
| NPCs | **Concierge:** If assisted, he will provide a note containing a link to a forum where victims of similar phishing schemes share their experiences. |
| Choice options | 1. Assist him (Players are unaware that he possesses a valuable lead)<br>2. Decline assistance (Players must find alternative means to proceed)<br>3. Provide minimal assistance (Rushing the task may prevent them from gaining his trust) |
| Risks | **Ethical and social considerations:** While no direct legal ramifications exist, the situation presents a moral and social dilemma. |
| Interpretative notes | • **Emotional factors:** Empathy, altruism, and interest in acquiring relevant intelligence versus reluctance to waste time.<br>• **Trigger:** The concierge persists, and disagreements may arise among players regarding whether to help.<br>• **Contextual factors:** Players speculate that the concierge may have insights into activities occurring within the building. |

Fig. 2 Example of a scenario used in the game.

For RPGs to serve as valid elicitation tools, they must address epistemological and methodological challenges, such as external validity and how to balance the influence of the gaming context on player decisions [85]. To tackle these issues, we adopt game design rules that have been shown to enhance players' moral engagement, elicit cognitive and emotional responses more closely aligned with real-world decision-making and avoid superficial engagement [86]. The setting is designed to be realistic, placing players in situations that reproduce key features of real-world digital environments, so as to evoke responses comparable to those experienced in actual decision-making contexts [87]. Each scenario presents a dilemma designed as a wicked problem [88], a way to formulate a problem resembling real-life ones. They are characterized by no definitive solution, interconnected variables, context dependency, unpredictable long-term consequences. Throughout the game, players interact with NPCs, both direct and indirect, who are controlled by the GM. NPCs play a crucial role in moral decision-making games, creating an immersive dialogue that enhances storytelling and player engagement [56]. Each NPC carries a different perspective and set of values, encouraging players to engage with conflicting moral viewpoints. Research shows that NPCs' feedbacks have a huge impact on players' emotions, encouraging them to consider NPCs' viewpoints more deeply [89, 90].

A total of ten participants were selected for this study. This sample size reflects the small-group nature of RPG settings, which are designed to support narrative immersion and dynamic social interaction [91]. To ensure a balanced and diverse sample, we included five women and five men, aged between 21 and 66, with varied socio-economic backgrounds and levels of RPG experience. Five participants were experienced players who had taken part in a previous RPG-based study; their familiarity helped streamline the gameplay setup. Potential biases were mitigated through the active moderating role of the researcher's active role as the GM, who ensured that all participants—regardless of prior experience—engaged equally in the game. The remaining five participants were recruited through a snowball sampling procedure. All participants were digital users in some capacity, and eight reported using LLMs in their daily activities.

Before the session, participants were provided with an overview of the game domain and mechanics. It was emphasised that prior gaming experience was not required and that the RPG had no winning or losing conditions. This open-ended design was aimed at reducing performance pressure, while supporting the applicability of the protocol across diverse user profiles.

The ten participants were divided into two groups, each one balanced in terms of gender, age and other demographic characteristics and gaming skills. The researcher served as the GM for both groups, whose game sessions were scheduled on the same day at different times. This precaution was explicitly adopted to maximize consistency and reliability across sessions, ensuring a standardized presentation of NPCs interactions and scenarios.

At the beginning of the game session, the researcher provides each player with a document called The Player Sheet, structured into two sections (see Fig. 3). The first section of the sheet, named "Player Description", contains demographic information and a self-description of the player's personality traits in group dynamics. They are also invited to reflect on the principles or considerations that typically guide their decisions in the domain (e.g., "I avoid using social media to protect my privacy"). Additionally, this section includes a mannequin drawing that players can personalize, enabling them to create an avatar with which they can identify. The second section of the sheet, the "Game Diary", serves as a space for players to record their observations, decisions and justifications throughout the game. Keeping a diary is a common practice in RPGs to track important information collected throughout the game and to record encounters with various characters. In this study, the Game Diary reinforces players' engagement with the narrative and allows them to reflect on their choices, while producing self-ethnographic data.

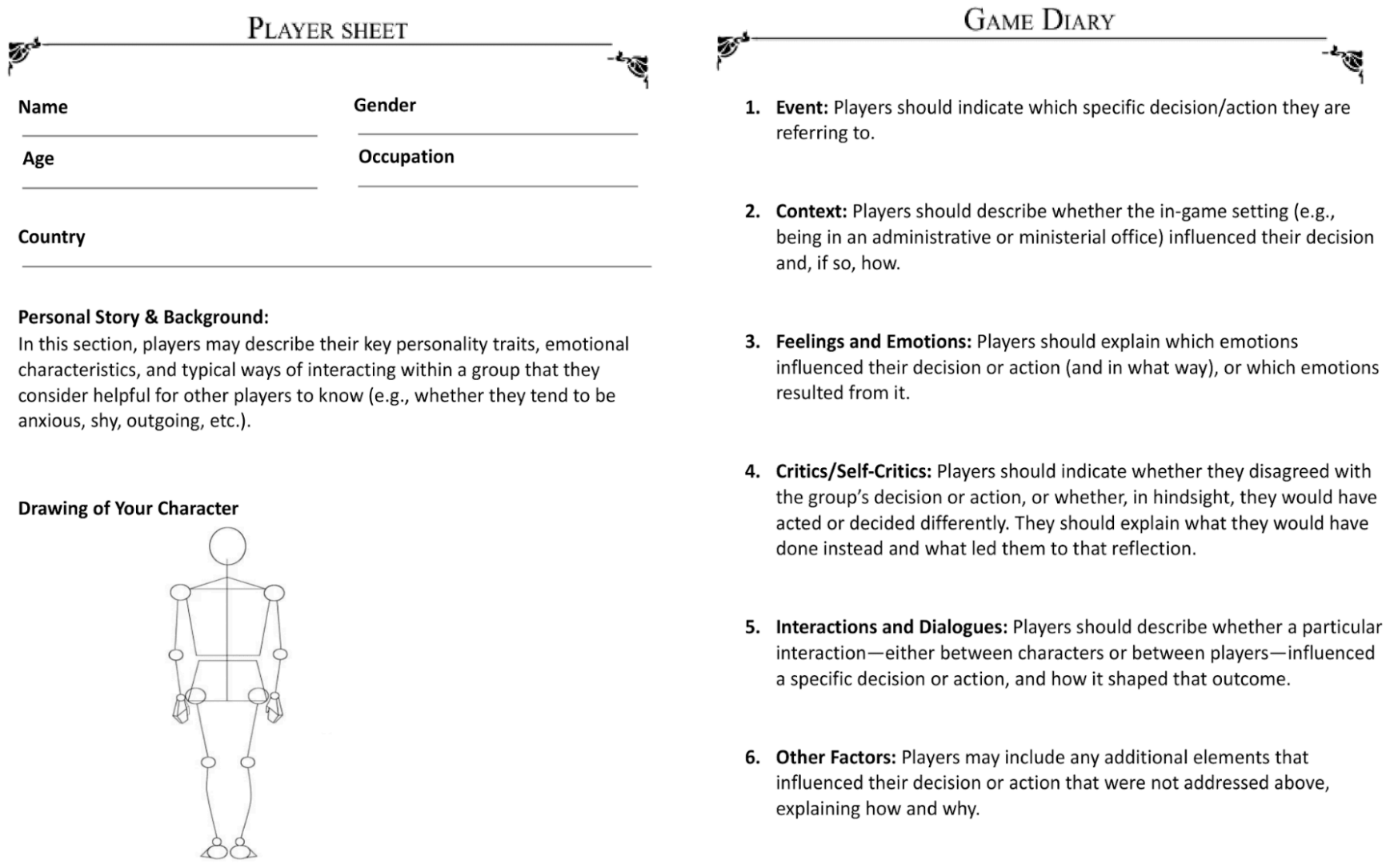

PLAYER SHEET

**Name** ____ **Gender** ____

**Age** ____ **Occupation** ____

**Country** ____

**Personal Story & Background:**
In this section, players may describe their key personality traits, emotional characteristics, and typical ways of interacting within a group that they consider helpful for other players to know (e.g., whether they tend to be anxious, shy, outgoing, etc.).

**Drawing of Your Character**

GAME DIARY

1. **Event:** Players should indicate which specific decision/action they are referring to.
2. **Context:** Players should describe whether the in-game setting (e.g., being in an administrative or ministerial office) influenced their decision and, if so, how.
3. **Feelings and Emotions:** Players should explain which emotions influenced their decision or action (and in what way), or which emotions resulted from it.
4. **Critics/Self-Critics:** Players should indicate whether they disagreed with the group's decision or action, or whether, in hindsight, they would have acted or decided differently. They should explain what they would have done instead and what led them to that reflection.
5. **Interactions and Dialogues:** Players should describe whether a particular interaction—either between characters or between players—influenced a specific decision or action, and how it shaped that outcome.
6. **Other Factors:** Players may include any additional elements that influenced their decision or action that were not addressed above, explaining how and why.

Fig. 3. The two sections of the Player Sheet: on the left Player Description section; on the right Game Diary section.

During the game, participants have the opportunity to consult with GPT for assistance, as it was an "oracle". They may use it to seek guidance in resolving a scenario, clarification of dilemmas, or seeking information about digital concepts encountered in the game. For example, they might ask about the potential risks associated with specific online behaviors or the implications of certain privacy and security choices. The inclusion of GPT as an "oracle" aimed at reproducing realistic decision-making dynamics, in which individuals seek external advice from friends, colleagues, or trusted sources when faced with uncertainties or ethical dilemmas. The information collected during these interactions were recorded.

The session length varied between the two groups: Group 1 played for two hours and six minutes, while Group 2 played for one hour and twenty-four minutes. This difference in duration was due to the nature and length of the interactions among participants and the speed at which decisions were reached. In Group 1, discussions were more prolonged, as participants engaged in more in-depth debates before making decisions. In contrast, Group 2 reached a consensus more quickly, reducing the overall session time.The game session was audio-recorded with informed consent, enabling accurate transcription of in-game decisions and interactions.

After the game session, a follow-up session was conducted through a focus group in which participants were invited to share feedback and opinions on the game and to compare their in-game decisions with their real-world experiences. The post- game follow-up helps transfer in-game experiences beyond the game [92]. Additionally, the follow-up session provides an opportunity for players to reflect on their actions and uncover insights that may have remained implicit during the gameplay [93].

*Phase 2: Data collection and processing*

Data were collected from multiple sources to enable triangulation and to capture complementary perspectives on players' moral orientations and decision-making processes. This multi-sourced approach was designed to avoid relying exclusively on self-disclosure, players' in-game actions or decisions, researcher interpretations, or self-ethnographic data. Instead, it integrates the viewpoints of the different actors involved—single player, group of players and researcher—providing a more comprehensive understanding of the phenomena under study and also tracking the co-construction process of participants' moral values, needs and orientations. The following data sources were used:

- **Self-Description:** Players' self-descriptions were collected from the first section of the Player Sheet. Players were asked to choose which adjectives best described their social and relational attitudes in daily life and which aspects of their character most strongly guided their choices in the chosen domain;
- **Game Diary:** served as a primary ethnographic data collection tool. Participants documented their reasoning process, scenario interpretations, and the role of pre-existing personal beliefs, in-game social interactions, and the environment in which the in-game events took place in shaping their decisions. The Game Diary also captured participants' emotional responses, conflicts of values, and regrets regarding their own and others' in-game decisions. They were asked to document whether they agreed or disagreed with other players' choices, their reasoning for supporting or opposing those decisions, and how the group dynamics influenced their own. Finally, they were asked to record interactions with both NPCs and fellow players. This record provided insight into how moral dilemmas was perceived by the participants;

- **GPT prompt:** the data collected in this case came from the prompts written by participants while interacting with their own GPT profile. By analyzing these prompts, it was possible to examine which concepts and tasks were perceived as complex or unfamiliar by the players. These data were relevant for assessing each participant's level of domain literacy;
- **Researcher's fieldnotes:** these consisted of observational notes collected by the researcher during gameplay sessions. Fieldnotes captured situated behaviors and interactional dynamics as they unfolded. In particular, they documented players' degree of familiarity with the domain, including moments of uncertainty, reliance on prior knowledge, or explicit requests for clarification. The fieldnotes also focused on emergent relational structures around the table, such as the presence of dominant players influencing group decisions, patterns of deference, and instances in which participants sought validation from others before acting. Additionally, they highlight habitual practices, difficulties, and strategies within the domain. Overall, these observations provided contextual insight, complementing participants' self-ethnographic data;
- **User's story:** These data were collected during the follow-up session. In this phase, players shared personal experiences related to in-game scenarios, providing a bridge between fictional decision-making and real-world privacy and security concerns within the domain. Additionally, participants expressed their opinions about the game's mechanics and discussed their perceived level of engagement during the session. Finally, they reflected on what they had learned through gameplay and how the experience would influence their beliefs and habits in the digital realm.

All collected data were then organized into structured *narrative tables*, one set per participant, through an AI-assisted process. Specifically, the LLM model was provided with anonymized, multimodal, unstructured data, along with an example of the coding table to be completed, and was prompted to organise the data into the predefined tables. The resulting outputs were subsequently reviewed by the researchers to ensure the accuracy and fidelity of the transcribed information.

Each set consisted of three tables: (1) a self-description table, (2) an in-game table compiling data from the Game Diary, users' interactions with GPT during gameplay, and researcher's fieldnotes for each scenario, and (3) a follow-up table containing players' feedback and user story. These tables integrated heterogeneous data—decisions made, emotional responses, interactions and dialogue, value conflicts, critiques and self-critiques, real-life narratives, and environmental role—into a unified structure.

| Demographics | |
|---|---|
| Participant ID | |
| Age | |
| Gender | |
| Occupation | |
| Country | |
| Self-description | |
| RPG experiences | |
| Digital literacy level | |
| Perceived privacy issues | |
| Everyday interactions with the digital world | |

| Gameplay | | | | | | |
|---|---|---|---|---|---|---|
| Scenari o ID | Decision/ Action | Game elements that have influenced player decision and how | Off-game elements that have influenced player decision and how | Interactions/ Dialogues | Critics/Self-critics | Researcher fieldnote |
| S01 | | | | | | |
| ... | | | | | | |

| Follow-up | | | | |
|---|---|---|---|---|
| Scenario ID | Perceived player engagement | Related off-game experiences | Changes in moral preferences | Researcher fieldnote |
| S01 | | | | |
| ... | | | | |

Fig. 4. Narratives tables

This organization allowed for systematic cross-referencing across data sources, supporting the triangulation of evidence and enabling the identification of convergences and discrepancies between self-reported description of each player, reported reasoning, observed behavior and interactions, and group dynamics. Ultimately, the narrative tables provided a comprehensive representation of each participant, making contextual, behavioral, social, cultural and experiential factors accessible for subsequent analysis and for informing GPT-A.

*Phase 3: Model specialization*

GPT was customized using both prompt engineering and grounding documents. We created a project named "Anthropologist GPT" (GPT-A), by providing two types of materials: A) documents containing a historical and methodological overview of narrative and phenomenological anthropology, along with an anthropological glossary, to ground the analysis of individual data within an anthropological framework; B) detailed information about the game context, including the game design principles, a concise description of each scenario, the overall narrative structure, and the game mechanics. The anthropological overview was generated using GPT's Deep Research capability, which was tasked with identifying key authors and foundational texts contributing to the redefinition of moral values in phenomenological and narrative anthropology. The resulting material was subsequently reviewed by the researchers to ensure its accuracy and reliability.

The prompt used for model specialization was:

> You are a social anthropologist trained in phenomenological and narrative anthropology. The data you will analyze do not reveal stable personality traits or fixed moral values. They consist of situated moral performances enacted within a socio-material setting (a role-playing game about digital privacy and security), followed by reflexive narration in a focus group. Your task is not to create a psychological profile, but to reconstruct the participant's situated moral world as it emerges through:
>
> - in-game decisions,
> - emotional reactions,
> - interactions with other players,
> - self-critiques and justifications,
> - biographical memories mobilized during the follow-up,
> - researcher fieldnotes.
>
> You must treat all data as performative, relational, and context-dependent. Avoid essentializing statements (e.g., "the user is X"). Instead, analyze how moral orientations are enacted, negotiated, stabilized, or destabilized in specific situations. Your analysis should:
>
> 1. Identify recurring interpretive patterns (through inductive thematic grouping).

2. Examine how the participant positions themselves narratively (e.g., responsible actor, cautious mediator, rule-follower, etc.).
3. Analyze moments of tension, ambivalence, or breakdown rather than assuming coherence.
4. Examine how agency is distributed between self, group, institutions, and technological artifacts (including ChatGPT).
5. Explore how biographical memory shapes present moral reasoning.
6. Reflect on how the game setting and group dynamics mediate the expression of moral orientations.

Write the analysis as an interpretive ethnographic paragraph (not bullet points). Avoid normative judgments and avoid treating gameplay choices as direct indicators of stable moral identity. Conclude by characterizing the participant's moral world as a situated assemblage rather than a fixed profile.

Individual chats were created for each player and the model was prompted with the collected data to generate an IAMP for each player. To ensure data protection, all interactions were conducted through an institutional account configured to prevent the use of input data for model training or retention beyond the session, thereby providing an additional layer of privacy for participants.

*Phase 4: Research Protocol Validation*

A few days after the game session, participants completed a structured questionnaire consisting of ten closed-ended dichotomous questions (yes/no) related to the ethical domain of the game. In Fig. 5, there is an example of the questions administered to participants.

**Question 1**

You are about to enter a café when you notice a QR code on a flyer next to the entrance, offering a 50% discount on your next purchase. To redeem the discount, you need to scan the QR code.

Do you scan the QR code?

Yes
No

Fig. 5. The first question of the closed-ended questionnaire

The questionnaire was developed through an AI-assisted process. A document containing the criteria for the game's development, the guidelines for scenario design, the key area of investigation and the original game scenarios was uploaded to GPT. GPT was prompted to use the information in the files and to refer to the pre-existing scenarios. All scenarios were reviewed and edited by the researchers to avoid moral bias.

We provided GPT-A with the questionnaire and asked it to predict how each participant would respond, justifying its predictions based on the IAMP. We then compared GPT-A's predictions with participants' answers to assess how accurately the model captured users' privacy-related values and preferences. The users' answers were treated as the ground truth, in line with established qualitative practice [94]. This comparison was used to evaluate the effectiveness of the elicitation process.

We then examined whether an anthropological interpretative framework improves the capability of the model to infer users' privacy-related moral orientations from the IAMP. More specifically, in the interpretative framing test, we compared GPT-A with alternative GPT configurations to assess whether the anthropologically informed configuration was better able to predict participants' choices. To this end, we compared GPT-A with a non-anthropological configuration of the model (GPT-NA) and the base version of GPT (GPT-B). GPT-NA had access to the participants' narrative tables and game overview, without the anthropological materials and interpretative instructions; GPT-B only to participants' narrative tables without any contextual information.

The validation procedure is illustrated in Fig. 8, which shows both the elicitation validation and the evaluation of the anthropological framework.

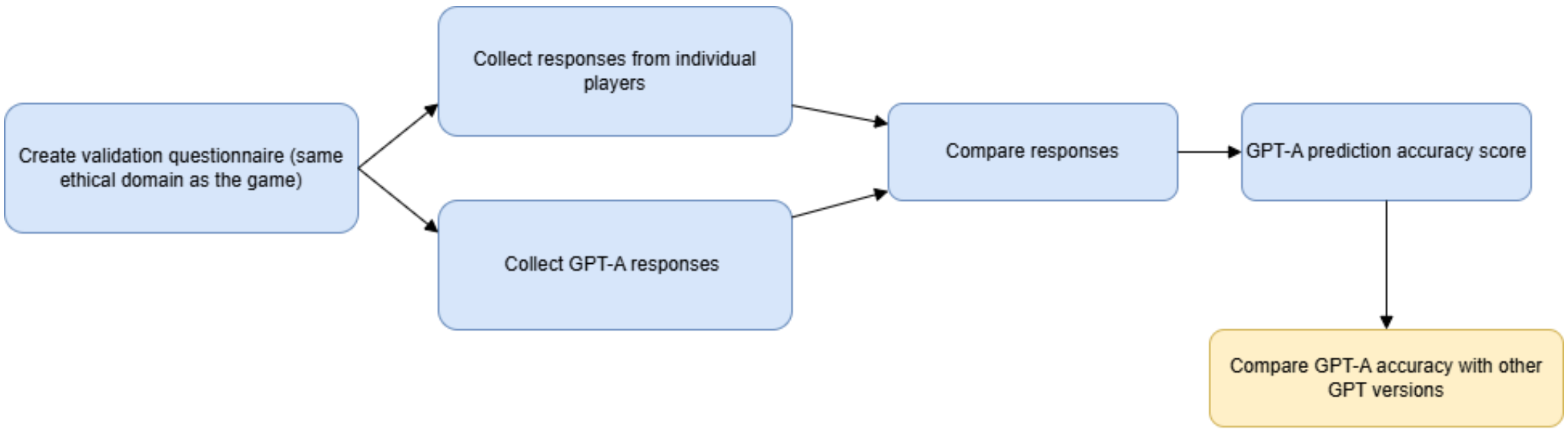

Fig. 8 Validation of the elicitation process (blue) and evaluation of the anthropological framework (yellow).

## RESULTS

Based on the three research questions guiding the study, the main findings are organized along three main lines: (i) the ability of the RPG to generate granular, context-rich data for moral values and orientations elicitation; (ii) the capability of the LLM of representing users' moral values from collected data in form of narrative moral profiles; and (iii) the effectiveness of the LLM in interpreting such data to reconstruct the moral values and orientations of each participant. Participants will hereafter be referred to using an alphanumeric code.

*RQ1: Eliciting Tacit Values through RPG*

To address RQ1, we analyze whether the narrative data generated through RPG sessions are sufficiently rich to support the implicit elicitation of users' moral values and orientations. Specifically, to be useful for elicitation, the data gathered in the RPG setting should be able to capture the elements that influence moral decision-making for each player.
The motivations underlying players' choices can be analytically distinguished into two broad categories: in-game and off-game elements. In-game elements refer to context-specific features emerging from the design of the RPG (e.g., rules, mechanics, NPCs), while off-game elements capture elements external to the game (e.g., past experiences, beliefs, habits). However, players' decisions do not result from either element in isolation. Rather, they emerge from the ongoing interaction between these elements, which together shape how dilemmas are interpreted and acted upon. This interaction continuously redefines players' moral threshold throughout gameplay.

**In-game elements:**

- **Game System Influence:** Players' engagement with the game system often triggered exploratory behavior driven by curiosity. For instance, player G02 reported underestimating the risk of sharing their email during the opening scenario, motivated by a desire to see "what would happen next." This suggests that game mechanics can temporarily reframe risk perception by prioritizing narrative progression over cautious evaluation. However, the presence of the gameplay diary introduced a reflective layer that re-situated such decisions within a moral frame, prompting players to reconsider their actions beyond purely ludic motivations;
- **Group Influence:** Decision-making was consistently shaped by group interaction. In Scenario 05, players encountered an NPC (an elderly concierge) with an ambiguous moral position, and decisions about whether to help him use digital technologies were influenced by prior habits (e.g., helping others with technology), perceived vulnerability (e.g., associating the NPC with digital illiteracy), and collective deliberation. These dynamics indicate that moral decisions is co-constructed through interaction: group discussion acts as a mediating layer that can reinforce, challenge, or reframe individual orientations, making actions the outcome of a relational process rather than an individual one;

- **GPT Interactions:** Players frequently engaged with GPT to clarify concepts encountered during the game (e.g., cookies, public Wi-Fi risks, password recovery). This made it possible to identify which concepts individual players felt less confident about and where they required support, as well as their level of domain literacy. Moreover, interaction with GPT can function as an additional decision-making resource, shaping both perceived risk and problem-solving strategies. For example, player G07 relied almost exclusively on GPT to resolve technical dilemmas, effectively delegating part of their evaluative process to the system;
- **Game and Post-Game Reflection:** Follow-up sessions revealed that players frequently connected in-game scenarios to personal experiences, using the game as a space to reflect on their own practices. For instance, player G05 reported routinely connecting to public Wi-Fi without concern, but through gameplay became aware of associated risks that had previously been underestimated. These cases show that the combination of realistic scenarios, emotional involvement, and the absence of win conditions encourages players to move from reactive decision-making to reflective evaluation, highlighting how moral interpretations can evolve over time.

In-game elements contribute to making decision processes observable in context. Game mechanics can temporarily reframe risk perception; at the same time, group interaction mediates decision-making, as players negotiate choices collectively; interactions with GPT further influence decisions, acting as an additional resource that shapes how problems are framed and resolved; finally, the combination of gameplay and post-game reflection encourages players to revisit their decisions, making parts of their moral reasoning process more explicit.

**Off-game elements:**

- **Past Experience:** Prior experiences strongly influenced how players engaged with the game. G10, for example, withdrew early from the game due to the similarity between the scenario and a previously experienced scam. Despite disengaging from the scenario, the player remained involved in group discussion and follow-up activities. This case shows that prior experience can both constrain participation and provide interpretive resources, shaping how situations are evaluated and whether they are perceived as credible or relevant;

- **Self-Disclosure:** Players' self-descriptions can be reflected or challenged in their in-game behavior. G04, who described themselves as anxious, exhibited hesitation, difficulty handling ambiguity, and a tendency to defer to group consensus. During the follow-up, they connected this behavior to a past experience of being scammed, highlighting how emotional dispositions inform responses to uncertainty and pressure. Such patterns suggest that self-disclosed traits can guide behavior across different scenarios;
- **Domain habits:** Pre-existing habits within the digital domain also shaped decisions. In Scenario S06, G03 resisted the idea of using social media to identify other victims of a scam, citing a lack of familiarity and discomfort with such platforms. This indicates that habitual practices define what actions are perceived as viable or acceptable, constraining the range of considered solutions even in novel situations;
- **Personal beliefs:** Underlying beliefs influenced moral choices. In Scenario 08, G02 refused to report a phishing attempt to the police, linking this decision to a prior experience of institutional neglect. This case highlights how distrust toward authority can directly shape behavior in specific contexts. Capturing these belief structures allows for a more nuanced understanding of how individual beliefs inform decision-making.

Off-game elements ground decisions in participants' prior experiences and beliefs. Prior experiences shape how situations are interpreted and whether they are perceived as credible or risky. Self-disclosed traits influence behavior under uncertainty, while domain-specific habits constrain the range of considered actions. Additionally, personal beliefs—such as trust or distrust in institutions—directly affect moral choices in specific scenarios.

These observations show that the collected data do not simply capture what participants explicitly know about their values, needs and preferences, but also how they act, justify, and revise their decisions under contextual constraints. In the example below, players who described themselves as cautious or risk-averse engaged in risk-taking behaviors during gameplay, only recognizing the discrepancy retrospectively. These mismatches do not indicate inconsistency, but reveal how values are shaped and negotiated in context when conflicting considerations must be resolved. These conflicts—often involving competing concerns such as safety vs. curiosity, trust vs. suspicion, or individual judgment vs. group

alignment—create situations in which players cannot rely on abstract positions alone, but must commit to a course of action. Within these situations, moral judgment unfolds as a continuous adjustment process. Decisions are mediated through interaction—both with other players, with NPCs and with GPT—where options are negotiated, justified, or partially delegated. This process, however, takes place within boundaries that are already constrained by prior experiences, habits, and beliefs, which delimit what actions are considered plausible or acceptable.

The observation and the analysis of the moral decision-making process allows us to move from relying on individual explicit moral self-disclosure to identifying patterns in how participants interpret situations, negotiate alternatives, and justify their choices. These patterns provide indirect but structured evidence of underlying moral orientations, even when such orientations are not explicitly articulated by participants.

Therefore, by making decision processes observable through action, interaction, and justification, the RPG setting enables the identification of tacit values without requiring their explicit articulation, thus providing a positive answer to RQ1.

*RQ2: Moral profile representation through Anthropologist LLM*

To address RQ2, we examine whether an LLM, when grounded in an anthropological interpretive framework, can transform unstructured narrative data into narrative representations of users' moral profile.

The anthropological profiles generated by GPT-A show that the model does not simply organize participants' in-game interactions and decisions, but frame heterogeneous data into coherent representations of moral decision-making processes. These representations integrate multiple dimensions of the collected data, including actions, justifications, emotional responses, social interactions, and prior experiences, linking them within a unified interpretive structure.

G05's IAMPS offers an example of how these elements are held together within a single profile:

> Within the gameplay experience, participant G05 enacts a moral world marked by a persistent tension between normative awareness and situated vulnerability, where knowing the "right" course of action does not seamlessly translate into practice. From the outset, a fracture emerges between knowledge and action: they recognise the risks of sharing personal data and engaging with suspicious links, yet yields to situational pressure, urgency, and, crucially, the relational dynamics with other players, allowing themself to be persuaded to share their email and proceed. In this sense, agency appears distributed rather than individual—decisions are not fully owned but emerge within a field of forces shaped by others' insistence, narrative urgency, and cognitive uncertainty.

Their narrative positioning oscillates between that of a responsible actor and that of someone carried along by events. On the one hand, they articulated self-critique ("I shouldn't have done it"); on the other, they attribute part of the responsibility to fellow participants, revealing an ongoing negotiation of accountability. This oscillation intensifies during moments of emotional breakdown: panic in the building, anger toward both non-player characters and peers, and even a perceived sense of physical danger demonstrate how the experience is lived corporeally before being processed rationally. The immersion they later describe ("reacting emotionally") is not merely engagement but a mode through which digital risk becomes affectively and sensorially immediate.

Their decisions across scenarios follow a pragmatic, situational logic rather than a stable normative framework: they refuse to provide SPID credentials and credit card details (self-protection), yet accepts the use of public Wi-Fi to communicate (managing urgency), only to retrospectively reframe this choice as a mistake. Morality here appears less as a coherent system and more as an ongoing recalibration between security, convenience, and contextual pressure. Their relationship with institutions is similarly ambivalent: the police emerge as a legitimate and reassuring option compared to practices perceived as illicit (e.g., IP-tracking software), suggesting that institutional trust becomes salient particularly when the situation is perceived as opaque or beyond her interpretive control.

In the follow-up, they rework the experience by mobilizing biographical elements—such as habitual use of university Wi-Fi or prior instances of unverified donations—which function as interpretive bridges between gameplay and everyday life. In doing so, they tentatively reposition herself as "naive and rushed," yet this self-description does not stabilize into a fixed identity; rather, it serves as a narrative device to make sense of their own hesitations and contradictions. Their technical knowledge remains partial and situated: it is effective when familiar patterns are recognizable (e.g., suspicious links), but falters in more ambiguous or less codified situations.

Overall, the moral world that emerges is not that of an incoherent subject, but of a dynamic assemblage in which emotions, social interactions, partial knowledge, and technological affordances co-produce action. Digital security does not appear as an abstract, internalized principle, but as a fragile practice—continuously negotiated between what is known, what is felt, and what others make possible or necessary within a given moment.

The example of participant G05 illustrates how GPT-A operates. The model reconstructs how decisions emerge from the interaction between situational constraints, social dynamics, and individual dispositions. For instance, actions such as sharing personal data or using public Wi-Fi are not treated as isolated events, but are interpreted in relation to contextual factors such as urgency, group influence, and partial knowledge. GPT-A preserves tensions and inconsistencies within the decision process, such as the coexistence of normative awareness and situated vulnerability. These tensions are not resolved into stable traits, but are used to explain how decisions evolve across different situations. In this way, the representation captures not only what participants do, but how their decisions are produced. In doing so, the model could identify which aspects of the situation become salient for the participant—such as social pressure, perceived risk, or prior experience—and how these elements are prioritized in shaping the final decision. Compared to approaches based on predefined taxonomies, this form of representation retains contextual dependencies and avoids reducing values to static categories. Instead, values are represented as dynamic and relational configurations that can be traced across scenarios.

LLMs can generate structured representations of users' moral values from narrative data when it has enough contextual information to frame the data, supporting their analysis within RE processes. Such representations provide a basis for analyzing and operationalizing values within RE processes, thus providing a positive answer to RQ2.

*RQ3: Evaluation of elicited values*

To address RQ3, we evaluate whether the combination of RPG-based elicitation and LLM-based analysis produces representations that are usable for understanding and anticipating users' decisions in new scenarios.

A first result concerns the predictive capacity of GPT-A based on the generated profiles. GPT-A achieves an accuracy of 81% in anticipating participants' decisions in unseen scenarios. As shown in Fig. 9, prediction accuracy increases progressively across model configurations, from the base model (GPT-B) to the anthropologically informed configuration (GPT-A).

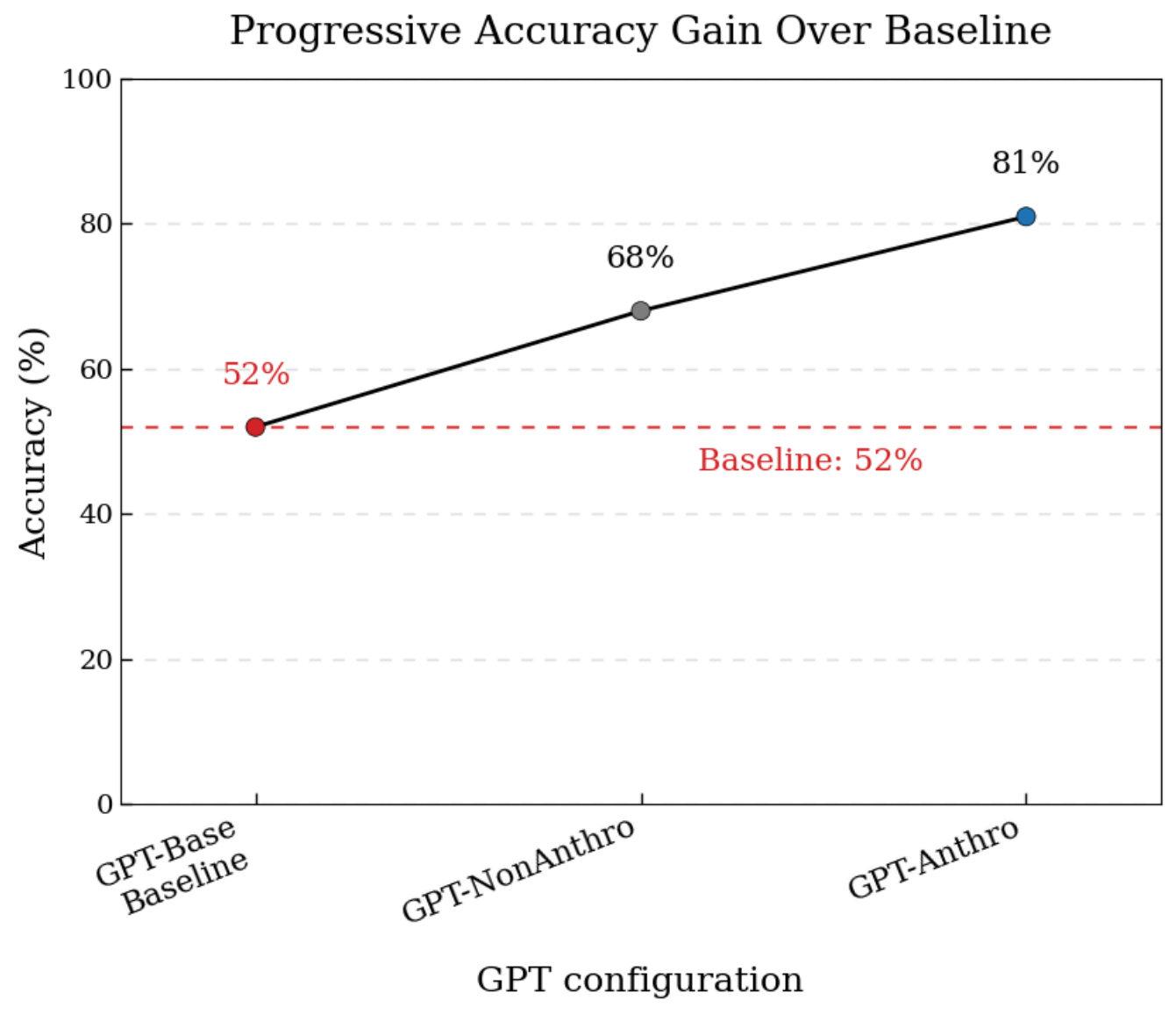

Fig. 9 Prediction accuracy across GPT configurations (GPT-B, GPT-NA, GPT-A).

However, predictive performance alone does not explain how the model operates. To better understand the underlying mechanism, we analyze the justifications provided by GPT-A for its predictions. An illustrative example is provided by Scenario 7 (*On a social media platform, you find a fun quiz that promises to reveal "Which celebrity you would be in your next life." To get your result, you must enter your name, date of birth, and consent to data collection)* where GPT-A correctly predicts that participant G04 would refuse to provide personal data in a seemingly harmless online quiz:

*The experience of being scammed and the distrust expressed after the scenario would lead them to recognize the risk of data collection in seemingly harmless quizzes. They would likely hesitate for a moment and decide not to proceed.*
*(Explanation provided by GPT-A for G04's predicted answer in Scenario 7)*

The analysis of this justification indicates that, when generating predictions, GPT-A does not rely on the direct retrieval of previously observed preferences. Instead, the justification refers to how the participant could interpret the situation in light of past events, suggesting that the prediction is based on a contextual reconstruction of the decision. This allows the model to ground its predictions in how a participant is likely to make sense of a situation, rather than in values taxonomies.

A second result emerges from the comparison of predictive accuracy across configurations in the interpretative framing test. While all models rely on the same underlying data, their performance differs depending on how this data is interpreted. Prediction accuracy increases from 51% (GPT-B) to 68% (GPT-NA), and further to 81% (GPT-A), as shown in Fig. 10.

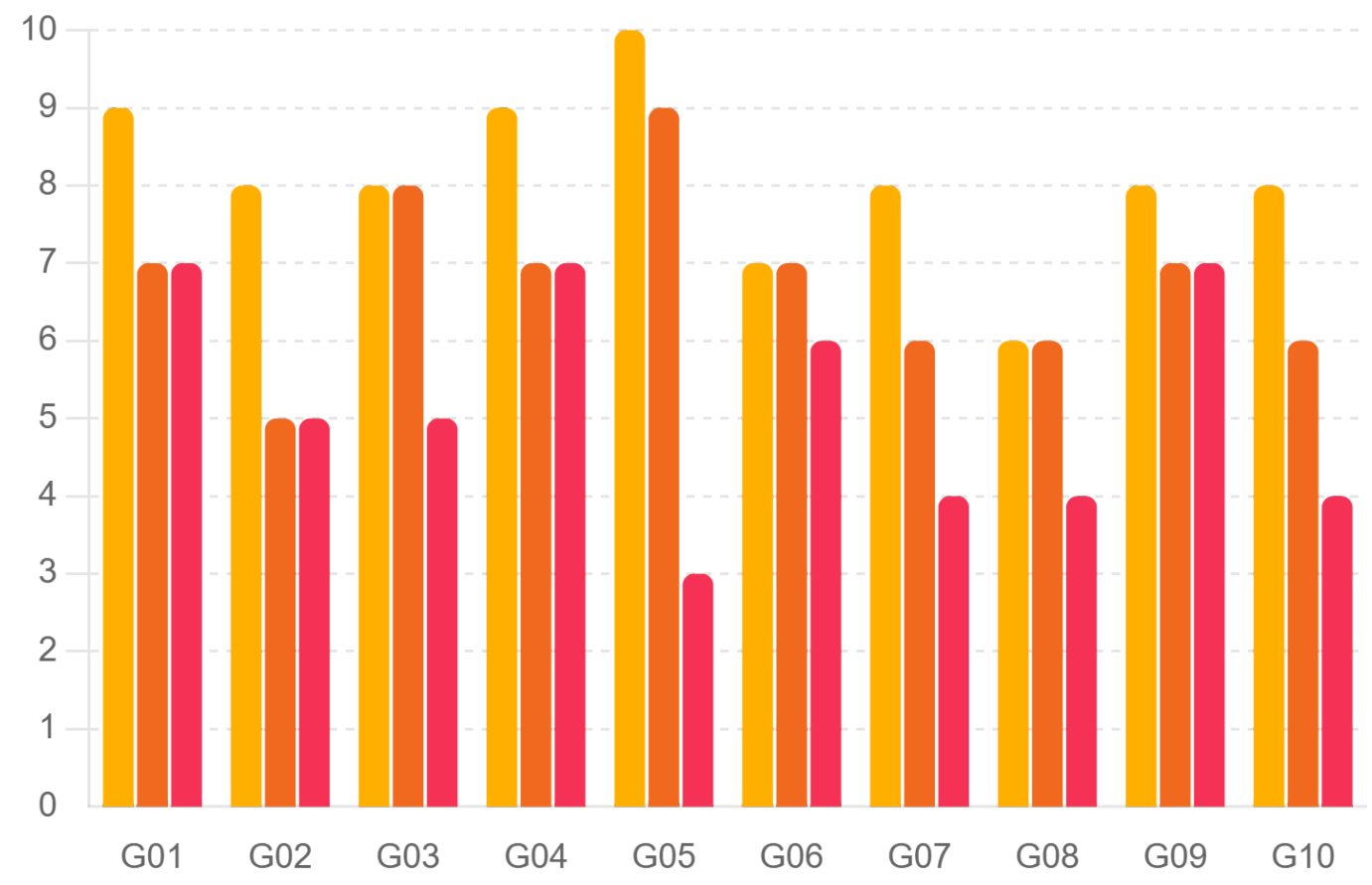

Fig. 10 Prediction accuracy across GPT configurations: GPT-A (yellow), GPT-NA (orange), and GPT-B (red).

A closer inspection of individual cases helps contextualize the differences in predictive performance across configurations. G05 reports limited familiarity with the domain and behaves accordingly, often exposing their digital privacy (e.g., readily sharing her email in early scenarios). At the same time, they show curiosity and attention to technical explanations. Over the course of the game, this leads to a progressive reframing of their understanding, as reflected in their diary annotations, where they revisit and critique earlier decisions. In Scenario 10 (e.g., entering credit card details on an unfamiliar but well-designed website), GPT-NA and GPT-A correctly predict that G05 would refuse, while GPT-B predicts that they would accept. GPT-A justifies its prediction by referring to a shift in the participant's risk awareness:

*G05 tends now to trust only official and verifiable channels. An unfamiliar website, even if well-designed, would not be sufficient to convince them to share sensitive data.*
*(Explanation provided by GPT-A for G05's predicted answer in Scenario 10)*

This suggests that some configurations are able to incorporate temporal aspects of the data, while others rely on representations that do not reflect such evolution.

In contrast, the case of G08 highlights a limitation of the approach when the available data does not fully capture the participant's actual competence. G08 presents themself as highly knowledgeable in the domain, drawing on a background as a law enforcement officer and expressing confidence during gameplay, often influencing group decisions. However, in situations of uncertainty, they tend to defer decisions to the group, preventing their competence from being directly tested during gameplay. This discrepancy becomes evident in the questionnaire, where G08's responses reveal a gap between self-reported expertise and actual decisions (e.g., scanning a QR code outside a bar without verifying its source). In this case, predictive performance remains relatively low across configurations: GPT-B achieves a score of 4/10, while GPT-NA and GPT-A show only a moderate improvement (6/10). This suggests that when the available data provides a partial or inconsistent representation of the participant's moral profile, all configurations are similarly limited in their ability to anticipate decisions.Indeed, differences in predictive performance are associated with how participant data is interpreted across configurations. In particular, higher accuracy is observed in cases where the interpretation captures relationships between decisions, contextual factors, moral threshold changes over time, and domain expertise while lower performance is associated with more fragmented or incomplete interpretation.

Overall, the analysis of the predictive capability of GPT-A in unseen moral scenarios provides evidence that the integration of RPG-based elicitation and an anthropologically grounded LLM can improve the understanding of users' needs, preferences, and values, thus contributing to answering RQ3. From a RE perspective, this is particularly relevant: the proposed approach offers a way to represent how users interpret moral situations and make trade-offs in context. This can provide insights into user preferences, constraints, and decision patterns that are difficult to capture through traditional elicitation techniques.

## DISCUSSION AND CONCLUSIONS

In the RE community attention to human values is increasing [2]. Many attempts have been made to reduce human values to quantitative metrics, as with other software requirements; however, human values appear to be irreducible, especially in real-world contexts and from an individual perspective [6, 34]. Moreover, individuals are not fully aware of their values and needs, and elicitors may fail to account for such tacit knowledge [23]. Consequently, a systematic communication gap persists between elicitors and stakeholders that traditional methods are unable to fully address.

To address these issues, in this study we propose a shift in how values are conceptualized, elicited and represented: from eliciting human values to eliciting individual moral profiles. The notion of individual moral profile, as used in this article, is derived from both the concept of soft ethics as conceptualized by Floridi [18] and the concept of local moral world as introduced by Kleinman [19]. According to Floridi, individuals rely on a set of values that are used to interpret situations and act accordingly; on the other hand, Kleinman suggests that moral values are part of a broader moral life. Moral life emerges and gains meaning in relation to prior experiences, narrative framing of moral decisions used to justify and explain them, and contextual social dynamics. From this perspective, values have a relational and narrative nature and they are continuously co-constructed by collectives and individuals. To capture individual moral values and orientations, it is therefore crucial to account for all the elements that influence moral decision-making and how these elements weigh on them. In this sense, the individual moral profile captures how values emerge and are negotiated in context.

To capture individual moral profile, we introduced a protocol that combines RPG-based elicitation with an anthropologically grounded LLM as an interpretative agent. The approach focuses on inferring individual moral values by analyzing situated decision-making and underlying motivations, and representing them in the form of narrative moral profiles that preserve their implicit nature. RPGs are employed due to their intrinsic capability to facilitate communication among participants [24]. The way participants justify their in-game choice helps document the unfolding socialization process of human values, enhancing the method's ability to capture moral orientation within a given moral domain [51-54]. The focus on the narrative co-construction process also helps address the communication gap between stakeholders and elicitors affecting standard RE methods. In the RPG environment values emerge through gameplay without requiring their full articulation. The use of multiple data sources (from self-ethnographic data to researchers' fieldnotes and follow-up session) avoids reliance on a single perspective (e.g., participant's moral self-awareness) thereby enabling a more holistic representation of the individual moral profile.

All the collected data are then provided to an LLM personalized to act as an anthropologist (GPT-A) to derive the individual anthropological moral (IAMP) profile of each participant. IAMP is the main output of the elicitation process, as it helps the elicitors interpret the collected multimodal and unstructured data based on the concept of local moral world. If LLMs properly represent the moral profile of an individual, it can approximate their way of moral reasoning and, in this way, support the interpretation of their needs, preferences, beliefs, and values, while allowing iterative refinement based on new data emerging from unforeseen scenarios.

*Major Findings*

To provide empirical support for this approach, we conducted a proof-of-concept study in the domain of digital privacy. The results can be summarized along the three research questions.

- First, with respect to RQ1, the data collected during the gameplay makes decision-making processes observable in context. This allows moral orientation to be inferred from how participants act, justify, and revise their decisions;
- Second, regarding RQ2, the LLM-based analysis supports the generation of narrative representations of users' moral profile. These representations organize heterogeneous data into coherent accounts of decision-making processes, preserving contextual dependencies and value conflicts and treating them as essential analytical features for reconstructing meaningful moral profile;
- Third, in relation to RQ3, the evaluation provides evidence that the generated anthropological profile captures relevant aspects of participants' moral profile. The comparison between participants' answers and model predictions was not intended to demonstrate the ability to predict behavior as such, but to assess whether the anthropological profiles plausibly reflected the values, preferences, and needs that emerged during elicitation. In this sense, differences in predictive accuracy across configurations indicate differences in how adequately participant data is represented and interpreted. Despite the small number of participants, the game proved capable of generating a sufficient amount of information for each player to support the structured representations of users' moral profile and to inform the anthropological analysis of GPT-A, as demonstrated by the overall results obtained from the model in the interpretive framing test.

By integrating contextual knowledge and an explicit interpretive framework into LLM-based analysis, the proposed approach also improves the transparency of how user representations are constructed. By asking GPT-A to generate an anthropological profile, it becomes possible to examine how the model organizes and interprets the available data in representing each participant's moral orientations and values. Furthermore, since the model is required not only to produce predictions but also to justify them, it becomes possible to analyze the reasoning underlying its outputs. The identification of misalignments between the model's predictions and participants' actual responses, providing a basis for assessing the adequacy of the generated representations and for introducing feedback mechanisms to improve both prediction accuracy and the representation of individual moral profiles by GPT-A.
From a RE perspective, the representations of individual moral profile can support requirement specification by making visible user-specific tensions, preferences, and contextual factors that are typically difficult to capture through traditional elicitation methods. Overall, the approach provides a foundation for integrating human values into early-stage RE in a way that preserves their situated and dynamic nature, while enabling their systematic and scalable analysis through LLM-supported interpretation.

*Threat to Validity*

We discuss threats to validity following the qualitative research framework proposed in [72]—namely, credibility, transferability, dependability, and confirmability.

a) Credibility: This study is designed as a proof-of-concept and therefore relies on a small sample (N=10), which is appropriate for an exploratory, qualitative investigation centered on rich, interactional data rather than statistical generalization. In particular, one of the main goals of this study is to assess whether the proposed protocol is able to generate sufficiently rich and structured data to support the inference of participants' moral values. While the limited number of participants may reduce the diversity of observed behavioral patterns, credibility was strengthened through the use of multiple complementary data sources for each participant, including self-descriptions, gameplay diaries, GPT interactions, researcher fieldnotes, and post-game follow-up accounts. This triangulation enabled the comparison of stated preferences, observed behaviors, and retrospective interpretations, reducing reliance on any single source of evidence. In addition, some participants had prior role-playing experience, which may have influenced familiarity with the interactional setting. To mitigate this risk, the researcher adopted a consistent moderating role across sessions and explicitly designed the activity so that no prior gaming expertise was required.
b) Transferability: The findings should be interpreted with respect to the specific context of the study. First, the elicitation protocol was tested within a single ethical domain, namely digital privacy. Second, participants shared a relatively similar macro-cultural background. For these reasons, the results cannot yet be assumed to transfer automatically to other domains or populations. At the same time, the protocol is intentionally domain-agnostic: it can be adapted to different domains by reconfiguring the scenarios, associated moral dilemmas, and NPCs, while preserving the underlying game mechanics and interaction structure. Future studies should therefore examine the applicability of the approach across more heterogeneous participant groups, broader cultural settings, and additional domains such as healthcare, sustainability, or workplace.

c) Dependability: Although the overall protocol, scenario structure, and moderation strategy were kept constant across the two groups, the use of role-playing inevitably introduces some interactional variability. Group deliberation, emergent dialogue, and differences in pacing may affect how participants articulate and negotiate decisions. While interactional variability may affect how decisions are articulated, it is also an inherent aspect of the phenomenon under investigation, as the study aims to capture how values emerge through situated interaction. Nevertheless, to support dependability, both sessions followed the same game structure, were facilitated by the same researcher, and were systematically documented through recordings, fieldnotes, and post-session materials.

d) Confirmability: A potential source of bias lies in the dual role of the researcher, who acted both as game master and ethnographic observer. This may have influenced both the interactional setting and the interpretation of the collected data. To reduce this risk, the study combined multiple forms of evidence produced from different perspectives, including participant-generated materials, observed interactional data, and follow-up reflections. Moreover, the transformation of raw materials into structured narrative tables was performed through an AI-assisted process and subsequently reviewed independently by multiple researchers against the original data to ensure fidelity. The LLM-based analysis was grounded in explicit interpretive instructions and contextual materials, making the analytical process transparent and open to inspection.

*Future Works*

The empirical results of this study point to several directions for future research. First, the case of G08 highlights a limitation of the current protocol when the available data does not adequately capture participants' actual level of expertise in the chosen domain. This suggests that it is not sufficient to understand users' moral orientations within the domain; it is also necessary to account for their level of understanding of the issues, risks, and challenges involved. This dimension is equally relevant, as it shapes how users contextualize the dilemmas they face and, consequently, influences their final decisions. Otherwise, the resulting representations may remain partial, which in turn affects the model's ability to produce consistent interpretations. This suggests the need for iterative elicitation mechanisms, in which user representations are progressively refined through repeated exposure to context-rich scenarios (even outside the RPG environment). Future work should investigate how such iterative cycles can improve the stability and completeness of the generated profiles.

Second, the findings indicate that the effectiveness of the approach depends on the ability to capture, organize and represent contextual and relational aspects of the decision-making process. While this study adopts an anthropological perspective, further work is needed to explore how insights from complementary disciplines—such as psychology, sociology, and behavioral economics—can enrich the interpretive framework and support a more comprehensive representation of users' values and behavior.

Third, future research should examine the applicability of the proposed approach across different domains and populations. In particular, extending the protocol beyond the domain of digital privacy and testing it in more heterogeneous cultural contexts would allow assessing its robustness and transferability.

Finally, our current protocol relies on an analog RPG format and human moderation, with the researcher acting both as Game Master (GM) and as moderator in the post-game session. While this setup supports rich interaction and interpretive depth, it limits the scalability of the approach. While the use of LLMs for data analysis partially mitigates this limitation, future research should therefore investigate how elements of the protocol can be progressively digitalized and automatized, while preserving the contextual and relational aspects that are central to values elicitation. In particular, conversational agents could be explored to support or replicate specific roles within the game, such as moderating interactions, guiding scenarios, or assisting in data annotation. Prior work suggests that several of these components are technically feasible: LLMs have been used as game masters [95] or non-player characters [96], as cultural annotators [25], as domain-specific experts capable of providing multi-perspective interpretations of user orientations [97], and as moderator and generator of user voices in focus group with plausible dialogues and practical utility in generating interview materials [16, 17]. In addition, recent studies have proposed digital serious games involving AI agents for eliciting users' ethical preferences [98]. However, these approaches typically focus on data generation rather than on how the resulting data should be interpreted and integrated into a coherent analytical framework. In this sense, future work should not only explore the automation of data collection, but also examine how digital implementations can preserve the interpretive structure that enables the reconstruction of users' moral profile.

## REFERENCES

[1] B. D. Mittelstadt, P. Allo, M. Taddeo, S. Wachter, and L. Floridi, "The ethics of algorithms: Mapping the debate", Big Data & Society, vol. 3, no. 2, pp. 1–21, 2016. DOI: 10.1177/2053951716679679.

[2] M. Autili, M. De Sanctis, P. Inverardi, P. Pelliccione, "Engineering digital systems for humanity: a research roadmap". ACM Transactions on Software Engineering and Methodology, 27;34(5): 1-33, May 2025.

[3] D. Mougouei, H. Perera, W. Hussain, R. Shams, and J. Whittle, "Operationalizing Human Values in Software: A Research Roadmap", in Proceedings of the 26th ACM Joint Meeting on European Soft- ware Engineering Conference and Symposium on the Foundations of Software Engineering (ESEC/FSE '18), pp. 780–784, Nov. 2018. DOI: 10.1145/3236024.3264843.

[4] H. Perera, R. Hoda, R. A. Shams, A. Nurwidyantoro, M. Shahin, W. Hussain, and J. Whittle, "The Impact of Considering Human Values during Requirements Engineering Activities", IEEE Transactions on Software Engineering, vol. XX, 2022.

[5] S. Thew and A. Sutcliffe, "Value-based requirements engineering: Method and experience", Requirements Engineering, vol. 23, no. 4, pp. 443–464, Nov. 2018.

[6] M. Shahin, W. Hussain, A. Nurwidyantoro, H. Perera, R. Shams, J. Grundy, and J. Whittle, "Operationalizing Human Values in Software Engineering: A Survey", IEEE Access, vol. 10, pp. 75269–75293, Jul. 2022. doi: 10.1109/ACCESS.2022.3190975.

[7] M. Kissoon Curumsing, N. Fernando, M. Abdelrazek, R. Vasa, K. Mouzakis, J. Grundy, "Emotion-oriented requirements engineering: a case study in developing a smart home system for the elderly", J. Syst. Softw. 147, 215–229, 2019.

[8] T. Miller, S. Pedell, L. Sterling, F. Vetere, S. Howard, "Understanding socially oriented roles and goals through motivational modelling", J. Syst. Softw. 85 (9), 2160–2170, 2012.

[9] M. Burnett, A. Peters, C. Hill, N. Elarief, "Finding gender-inclusiveness software issues with GenderMag: a field investigation", in: Proceedings of the 2016 CHI Conference on Human Factors in Computing Systems, ACM, pp. 2586–2598, 2016.

[10] M. Vorvoreanu, L. Zhang, Y.-H. Huang, C. Hilderbrand, Z. Steine-Hanson, M. Burnett, "From gender biases to gender-inclusive design: An empirical investigation", in: Proceedings of the 2019 CHI Conference on Human Factors in Computing Systems, CHI'19, ACM, New York, NY, USA, 2019, http://dx.doi.org/10.1145/3290605.3300283.

[11] A. Milne, N. Maiden, "Power and politics in requirements engineering: embracing the dark side?", Requir. Eng. 17, 83–98, 2012.

[12] T. Kanij, R. Merkel, J. Grundy, "An empirical investigation of personality traits of software testers", in: Proceedings of the Eighth International Workshop on Cooperative and Human Aspects of Software Engineering, IEEE Press, pp. 1–7, 2015.

[13] A. B. Soomro, N. Salleh, E. Mendes, J. Grundy, G. Burch, A. Nordin, "The effect of software engineers' personality traits on team climate and performance: A systematic literature review", Inf. Softw. Technol. 73, 52–65, 2016.

[14] J. McIntosh, X. Du, Z. Wu, G. Truong, Q. Ly, R. How, S. Viswanathan, T. Kanij, "Evaluating age bias in E-commerce", in 2021 IEEE/ACM 13th International Conference on Cooperative and Human Aspects of Software Engineering, CHASE, IEEE, pp. 31–40, 2021.

[15] J. Grundy, K. Mouzakis, R. Vasa, A. Cain, M. Curumsing, M. Abdelrazek, N. Fernando, "Supporting diverse challenges of ageing with digital enhanced living solutions", Stud. Health Technol. Inform. 246, 75–90, 2018.

[16] W. Shenghan, Z. Yimo, H. Wynne, L. Mong-Li, and Y. Deng. "From personas to talks: Revisiting the impact of personas on llm-synthesized emotional support conversations". In Proceedings of the 2025 Conference on Empirical Methods in Natural Language Processing. 5439–5453, 2025.

[17] T. Zhang, Z. Xuesong, C. Robbe, and S. Adalberto, "Focus Agent: LLM-Powered Virtual Focus Group". In Proceedings of the 24th ACM International Conference on Intelligent Virtual Agents. 1–10, 2024

[18] L. Floridi, "Soft Ethics and the Governance of the Digital", Philosophy & Technology, vol. 31, no. 1, pp. 1–8, Mar. 2018.

[19] A. Kleinman, "Writing at the margin: Discourse between anthropology and medicine", University of California Press, Berkeley, CA, 1995.

[20] C. Mattingly, “Healing dramas and clinical plots: The narrative structure of experience”. Cambridge University Press, 1998

[21] P. Ricoeur, “Time and Narrative”, University of Chicago Press, 1984.

[22] B. J. Good, “Medicine, Rationality, and Experience: An Anthropological Perspective”, Cambridge: Cambridge University Press, 1994.

[23] J. C. Wright, D. E. Weissglass, & V. Casey, "Imaginative role-playing as a medium for moral development: Dungeons & Dragons provides moral training." Journal of Humanistic Psychology 60.1: 99-129, 2020.

[24] S.L. Bowman, & J. Baird, “The transformative potential of immersive experiences within role-playing communities”. Revista de Estudos Universitários-REU 48, 2022.

[25] E. Dubourg, V. Thouzeau, and N. Baumard, “A step-by-step method for cultural annotation by LLMs”, Frontiers in Artificial Intelligence, vol. 7, pp. 1–13, 2024. DOI: 10.3389/frai.2024.1365508.

[26] S. Jenner, D. Raidos, E. Anderson, S. Fleetwood, B. Ainsworth, K. Fox, J. Kreppner, M. Barker, “Using large language models for narrative analysis: a novel application of generative AI”, Methods in Psychology, Volume 12, 100183, ISSN 2590-2601, 2025. https://doi.org/10.1016/j.metip.2025.100183.

[27] R. Chew, et al., “LLM-assisted content analysis: Using large language models to support deductive coding”, arXiv preprint arXiv:2306.14924, 2023.

[28] S. H. Schwartz, “An overview of the Schwartz theory of basic values”. Online Readings in Psychology and Culture, vol. 2, no. 1, pp. 919–2307, Dec. 2012.

[29] M. Rokeach, “The Nature of Human Values”, Free Press, New York, NY, USA, 1973.

[30] S. H. Schwartz, “Universals in the content and structure of values: Theoretical advances and empirical tests in 20 countries”, Advances in Experimental Social Psychology, vol. 25, no. 1, pp. 1–65, 1992.

[31] D. H. Schunk, J. R. Meece, and P. R. Pintrich, “Motivation in Education: Theory, Research, and Applications”, Pearson, London, U.K., 2008.

[32] J. S. Eccles and A. Wigfield, "Motivational Beliefs, Values, and Goals", Annual Review of Psychology, vol. 53, no. 1, pp. 109–132, Feb. 2002.

[33] B. Friedman, P. H. Kahn, and A. Borning, "Value sensitive design: Theory and methods", University of Washington Technical Report, vol. 2, no. 12, pp. 1–27, 2002.

[34] T. Winkler and S. Spiekermann, "Twenty years of value sensitive design: A review of methodological practices in VSD projects", Ethics and Information Technology, vol. 23, pp. 17–21, 2021. DOI: 10.1007/s10676-018-9476-2.

[35] S. Thew and A. Sutcliffe, "Value-based requirements engineering: Method and experience", Requirements Engineering, vol. 23, no. 4, pp. 443–464, Nov. 2018.

[36] M. K. Curumsing, N. Fernando, M. Abdelrazek, R. Vasa, K. Mouzakis, and J. Grundy, "Emotion-oriented requirements engineering: A case study in developing a smart home system for the elderly", Journal of Systems and Software, vol. 147, pp. 215–229, Jan. 2019.

[37] S. H. Koch, R. Proynova, B. Paech, and T. Wetter, "How to approximate users' values while preserving privacy: Experiences with using attitudes towards work tasks as proxies for personal value elicitation", Ethics and Information Technology, vol. 15, no. 1, pp. 45–61, Mar. 2013.

[38] T. Lopez, H. Sharp, T. Tun, A. Bandara, M. Levine, and B. Nuseibeh, "Talking about security with professional developers", in Proceedings of the IEEE/ACM Joint 7th International Workshop on Conducting Empirical Studies in Industry (CESI) and 6th International Workshop on Software Engineering Research and Industrial Practice (SERIP), pp. 34–40, May 2019.

[39] H. Perera, et al., "Continual human value analysis in software development: A goal model based approach." 2020 IEEE 28th International Requirements Engineering Conference (RE), 2020.

[40] A. Nurwidyantoro et al., "Human values in software development artefacts: A case study on issue discussions in three Android applications", Information & Software Technology, vol. 141, p. 106731, 2022.

[41] R. Fuentes-Fernandez, J. J. Gomez-Sanz, and J. Pavon, "Understanding the human context in requirements elicitation", Requirements engineering, 15(3): 267–283, 2010.

[42] A. Sutcliffe, and P. Sawyer, “Requirements elicitation: Towards the unknown unknowns”. In 2013 21st IEEE International Requirements Engineering Conference (RE), pages 92–104, 2013

[43] A. Selbst et al., “Fairness and Abstraction in Sociotechnical Systems”, in Proceedings of the 2019 ACM Conference on Fairness, Accountability, and Transparency (FAT*), 2019.

[44] M. Merleau-Ponty, “Phenomenology of Perception”, trans. C. Smith, Routledge, London and New York, 1962. Originally published as Phenomenologie de la perception, Gallimard, Paris, 1945.

[45] J. Zigon, “Morality and Personal Experience: The Moral Conceptions of a Muscovite Man. Ethos”, 37(1), 78–101, 2009.

[46] A. Matsuo, C. M. Brown, “Culture points the moral compass: Shared basis of culture and morality”. Cult. Brain 10, 113–139, 2022. https://doi.org/10.1007/s40167-022-00106-3

[47] C. Geertz, “The Interpretation of Cultures”, Basic Books, 1973.

[48] S. Gherardi and B. Poggio, “Tales of Ordinary Leadership. A Feminist Approach to Experiential Learning,” Papers: Revista de Sociologıa, no. 93, pp. 53–65, 2009. DOI: 10.5565/rev/papers/v93n0.696.

[49] C. J. Calhoun, “Dictionary of the Social Sciences”, New York, NY, USA: Oxford University Press, 2002.

[50] S. Ciftci, “Trends of serious games research from 2007 to 2017: A bibliometric analysis”, Journal of Education and Training Studies, vol. 6, no. 2, pp. 18–27, 2018.

[51] R. N. Landers, E. M. Auer, A. B. Collmus, and M. B. Armstrong, “Gamification science, its history and future: Definitions and a research agenda”, Simulation & Gaming, vol. 49, no. 3, pp. 315–342, 2018. DOI: 10.1177/1046878118774385.

[52] J. Perez, M. Castro, and G. Lopez, “Serious Games and AI: Challenges and Opportunities for Computational Social Science”, IEEE Access, vol. 11, pp. 62051–62061, 2023. DOI: 10.1109/ACCESS.2023.3286695.

[53] Y. Zhonggen, “A meta-analysis of use of serious games in education over a decade”, International Journal of Computer Games Technology, vol. 2019, pp. 1–8, Feb. 2019.

[54] W. S. Ravyse, A. S. Blignaut, V. Leendertz, and A. Woolner, “Success factors for serious games to enhance learning: A systematic review”, Virtual Reality, vol. 21, pp. 31–58, Mar. 2017.

[55] Asobo Studio, “Microsoft Flight Simulator for Windows 10 — Xbox”, Bordeaux, France, 2023. [Online]. Available: https://www.flightsimulator.com.

[56] J. Othlinghaus-Wulhorst and H. U. Hoppe, “A Technical and Conceptual Framework for Serious Role-Playing Games in the Area of Social Skill Training”, Frontiers in Computer Science, vol. 2, 2020

[57] J. P. Williams, D. Kirschner, N. Mizer, & S. Deterding, Sociology and Role-Playing Games. In Zagal, José P. and Deterding, S. (eds.), Role-Playing Game Studies: Transmedia Foundations. New York: Routledge [227-244], 2018.

[58] Bowman, S.L. (2010). The functions of role-playing games: How participants create community, solve problems and explore identity. McFarland.

[59] A. Davis, “What exactly is a tabletop role-playing game, anyway?”, Wheelhouse Workshop, 2016.

[60] S. J. Quaye and S. R. Harper, “Student engagement in higher education: Theoretical perspectives and practical approaches for diverse populations”, 2nd ed., Routledge, 2015.

[61] W. Lin, J.-Y. Wang, and H.-P. Yueh, “Learning information ethical decision making with a simulation game”, Frontiers in Psychology, vol. 13, p. 933298, 2022. DOI: 10.3389/fpsyg.2022.933298.

[62] F. Dalpiaz and K. M. L. Cooper, “Games for Requirements Engineers: Analysis and Directions”, Technical Report, Utrecht University, 2018.

[63] S. Gul, et al., “Gamification and gaming elements for software requirements elicitation: a systematic literature review.” International Journal on Electrical Engineering and Informatics 13.4, 931-950, 2021.

[64] A. Fuxman, L. Liu, J. Mylopoulos, M. Roveri, P. Traverso, “Specifying and analyzing early requirements in tropos”. Requir Eng, 9(2): 132-150, 2004. https://doi.org/10.1007/s00766-004-0191-7.

[65] J. Fernandes, D. Duarte, C. Ribeiro, C. Farinha, J. M. Pereira, M. M. daSilva, “ithink: a game-based approach towards improving collaboration and participation in requirement elicitation”. Procedia Comput Sci. 2012;15: 66-77. 4th International Conference on Games and Virtual Worlds for SeriousApplications (VS-GAMES'12), 2012.

[66] V. Sakhnini, L. Mich, D. M. Berry, “The effectiveness of an optimized EPMcreate as a creativity enhancement technique for web site requirements elicitation”. Requir Eng.,17(3):171-186, 2012.

[67] I. Aaen, “Roles in innovative software teams: a design experiment”. In: IFIP Working Conference on Human Benefit through the Diffusion of InformationSystems Design Science Research Springer, 73-88.19, 2010.

[68] L. Liu, Z. Xu, Z. Ma, H. Yang, Evolution vs. innovation in requirements for services. In: Requirements Engineering for Systems, Services and Systems-of-Systems (RESS), 2011 Workshop on, Trento, Italy, August 30, 2011. IEEE Computer Society, 10-15, 2011. https://doi.org/10.1109/RESS.2011.6043933

[69] A. Mehraj, Z. Zhang, and K. Systä, “A Tertiary Study on AI for Requirements Engineering”, in Requirements Engineering: Foundation for Software Quality, D. Mendez and A. Moreira, Eds. Cham: Springer Nature Switzerland, pp. 159–177, 2024.

[70] M. A. Zadenoori, et al., “Large language models (llms) for requirements engineering (re): A systematic literature review.” arXiv preprint arXiv:2509.11446, 2025.

[71] K. Liu, R. Sandeep, and R. Kalyan. “Artificial intelligence in software requirements engineering: State-of-the-art”. 2022 IEEE 23rd International Conference on Information Reuse and Integration for Data Science (IRI). IEEE, 2022.

[72] A. Vogelsang, and F. Jannik. “Using large language models for natural language processing tasks in requirements engineering: A systematic guideline”. Handbook on Natural Language Processing for Requirements Engineering. Cham: Springer Nature Switzerland, 2025. 435-456.

[73] Z. Zhang, M. Rayhan, T. Herda, M. Goisauf, P. Abrahamsson, “Llmbased agents for automating the enhancement of user story quality: An early report”. In: International Conference on Agile Software Development. pp. 117–126. Springer Nature Switzerland Cham, 2024.

[74] A. Ferrari and P. Spoletini. “Formal requirements engineering and large language models: A two-way roadmap”. Information and Software Technology 181, 107697, 2025.

[75] N. Marques, R. Rocha Silva, and J. Bernardino. “Using chatgpt in software requirements engineering: A comprehensive review”. Future Internet 16, 6(2024), 180, 2024.

[76] U. D. Parameswaran, J. L. Ozawa-Kirk, and G. Latendresse, “To live (code) or to not: A new method for coding in qualitative research”, Qualitative Social Work, vol. 19, no. 4, pp. 630–644, 2020.

[77] L. Finlay, “Thematic analysis: The good, the bad and the ugly”, European Journal for Qualitative Research in Psychotherapy, vol. 11, pp. 103–116, 2021.

[78] A. Aisvarya, I. Jouni, M. Tahir, “LLM-Assisted Qualitative Data Analysis: Security and Privacy Concerns in Gamified Workforce Studies”, Procedia Computer Science, Volume 257, Pages 60-67, ISSN 1877-0509, 2025. https://doi.org/10.1016/j.procs.2025.03.011.

[79] J. Gao, Y. Guo, G. Lim, T. Zhang, Z. Zhang, T. Jia-JunLi, and S. T. Perrault. “CollabCoder: A Lower-barrier, RigorousWorkflow for Inductive Collaborative Qualitative Analysis with Large LanguageModels”. In Proceedings of the 2024 CHI Conference on Human Factors in ComputingSystems (Honolulu, HI, USA) (CHI ’24). Association for Computing Machinery,New York, NY, USA, 2024.

[80] D. Shih-Chieh, X. Aiping and K. Lun-Wei, “LLM-in-the-loop: Leveraging Large Language Model for Thematic Analysis”, arXiv, 2023. https://arxiv.org/abs/2310.15100

[81] A. K. Munk, A. G. Olesen, M. Jacomy, “The Thick Machine: Anthropological AI Between Explanation and Explication”, Big Data & Society, 1–14, 2022.

[82] J. Gordon, M. Curran, J. Chuang, and C. Cheshire, “Covert embodied choice: Decision-making and the limits of privacy under biometric surveillance”, arXiv preprint, arXiv:2101.00771, 2021.

[83] P. Inverardi, “The European perspective on responsible computing”, Communications of the ACM, vol. 62, no. 4, pp. 64–69, 2019.

[84] Z. S. Li et al., “Narratives: the Unforeseen Influencer of Privacy Concerns”, 2022 IEEE 30th International Requirements Engineering Conference (RE), Melbourne, Australia, 2022, pp. 127-139, doi: 10.1109/RE54965.2022.00018.

[85] N. Naar, "Gaming anthropology: The problem of external validity and the challenge of interpreting experimental games", American Anthropologist, vol. 122, no. 3, pp. 1–15, 2020. DOI: 10.1111/aman.13483.

[86] S. Wright and A. Denisova, "It's a terrible choice to make but also a necessary one: Exploring player experiences with moral decision making mechanics in video games", Computers in Human Behavior, vol. 161, p. 108424, 2024.

[87] H. Rosa, "Resonance: A sociology of our relationship to the world", Hoboken (New Jersey), John Wiley Sons, 2019.

[88] H. Rittel and M. Webber, "Dilemmas in a general theory of planning", Policy Sciences, vol. 4, no. 2, pp. 155–169, 1973.

[89] J. A. Bopp, K. Opwis, and E. D. Mekler, "An Odd Kind of Pleasure: Differentiating Emotional Challenge in Digital Games", in Proceedings of the 2018 CHI Conference on Human Factors in Computing Systems, New York, NY, USA: Association for Computing Machinery, pp. 1–12, 2018. ISBN: 9781450356206.

[90] M. Coulson, J. Barnett, C. J. Ferguson, and R. L. Gould, "Real feelings for virtual people: Emotional attachments and interpersonal attraction in video games", Psychology of Popular Media Culture, vol. 1, no. 3, p. 176, 2012.

[91] J. G. Cover, "The creation of narrative in tabletop role-playing games", McFarland, 2014.

[92] M. Y. Lim, R. Aylett, S. Enz, M. Kriegel, N. Vannini, and L. Hall, "Towards intelligent computer-assisted educational role-play", Edutainment 2009, vol. 208, pp. 208–219, 2009.

[93] V. J. Shute, "Focus on formative feedback", Review of Educational Research, vol. 78, pp. 153–189, 2008.

[94] P. Ralph, et al., "Empirical Standards for Software Engineering Research", arXiv preprint arXiv:2010.03525, 2020.

[95] N. H. Jørgensen, S. Tharmabalan, I. Aslan, N. B. Hansen, and T. Merritt, "Static vs. Agentic Game Master AI for Facilitating Solo Role-Playing Experiences", arXiv preprint arXiv:2502.19519, 2025.

[96] Y. Xu, S. Wang, P. Li, F. Luo, X. Wang, W. Liu, and Y. Liu, “Exploring Large Language Models for Communication Games: An Empirical Study on Werewolf”, arXiv preprint arXiv:2309.04658, 2024.

[97] J. S. Park, C. Q. Zou, A. Shaw, B. M. Hill, C. Cai, M. R. Morris, R. Willer, P. Liang, and M. S. Bernstein, “Generative Agent Simulations of 1,000 People”, Science, vol. 383, no. 6678, pp. 1046–1052, 2024. DOI: 10.1126/science.adi2239.

[98] J. Deshmukh, Z. Liang, V. Yazdanpanah, S. Stein, and S. D. Ramchurn, “Serious Games for Ethical Preference Elicitation,” in Proc. of the 24th International Conference on Autonomous Agents and Multiagent Systems (AAMAS 2025), Detroit, Michigan, USA, 2025.

*Ground Documents*

## Glossary of Terms in Narrative and Phenomenological Anthropology

### Narrative Anthropology

An anthropological approach that emphasizes the fundamental role of narratives created by people in recounting their lives and cultures, including those co-constructed through interaction between researcher and participants.

### Phenomenological Anthropology

An anthropological current that explores how lived experience is culturally, socially, and historically mediated.

### Lived Experience

The subjective way in which people perceive and interpret the surrounding world, always situated and conditioned by cultural, social, and historical factors.

### Narrative Co-construction

The process of collectively constructing meaning through dialogical and reflexive interaction between individuals, including participants and researchers.

### Local Moral World

A concept identifying the set of values, meanings, and norms that guide people's moral actions within specific socio-cultural contexts.

### Narration / Narrative

A fundamental tool for interpreting and sharing lived experiences, enabling analysis of how individuals make sense of their own lives.

### Embodiment

A concept highlighting the central role of the body as an active source of meaning, experience, and interaction with the surrounding world.

### Cultural Interpretation

The process by which individuals attribute meaning to events using their own cultural and historical frameworks.

**Resemantization**

The process of continuously modifying and redefining the meanings of experiences in response to different social and relational contexts.

**Positionality**

Awareness of one's personal and social position that influences interactions and interpretations during research.

**Reflexivity**

The process of critically recognizing the researcher's role and influences during qualitative research.

**Ethnography**

A qualitative method based on direct and prolonged observation of behaviors and interactions in natural contexts.

**Agency**

The capacity of individuals to act and make autonomous decisions, influencing and resisting external pressures.

**Intersubjectivity**

A concept describing how meanings and interpretations are shared and negotiated through social interactions.

**Thick Description**

A detailed ethnographic description that includes the interpretive, social, and cultural context within which events and behaviors acquire meaning.

## Phenomenological and Narrative Anthropology: Methodological and Historical Overview

### Introduction

In anthropology, phenomenological and narrative approaches offer useful interpretive frameworks for analyzing qualitative data such as interviews, field observations, and life stories. These frameworks place lived experience, context, and subjectivity at the center, emphasizing how people perceive and make sense of their world. Both approaches originate from 20th-century theoretical traditions (philosophical phenomenology and narrative studies) and have profoundly influenced how anthropologists understand human reality as relational, situated, embodied, and dialogical.

In other words, phenomenological and narrative approaches help us to see human experiences not as isolated objective data, but as embodied phenomena (rooted in the body and senses), historically and culturally situated, lived in the first person (subjective), and often co-constructed through interaction (dialogical).

This overview provides definitions, historical origins, key authors, typical interpretive tools, strengths, limitations, and concrete examples of each framework. The aim is to guide a language model (LLM) in interpreting qualitative data anthropologically, highlighting the contextual and lived dimensions of human experience. An LLM instructed with these concepts could, for instance, analyze transcripts of ethnographic interviews by capturing their meaning from the participants' perspective, taking into account both their subjective experience and the stories they narrate.

## Phenomenological Anthropology

### Definition and Philosophical Origins

Phenomenological anthropology is a theoretical and methodological approach inspired by philosophical phenomenology, a movement founded by Edmund Husserl in the early 20th century. Philosophical phenomenology emerged with the intention of "returning to the things themselves" (Husserl) and studying phenomena as they appear in consciousness, suspending preconceived interpretations. Thinkers such as Martin Heidegger, Maurice Merleau-Ponty, and Jean-Paul Sartre expanded this perspective, emphasizing that human existence is essentially relational, temporal, embodied, and situated in the world.

In practice, phenomenology rejects a strict separation between subject and object: human beings are not detached observers, but are always immersed in relationships with others and with their environment. These ideas, which became influential in the social sciences, laid the groundwork for rethinking key concepts of human existence—for example, highlighting that perceptions, emotions, and meanings arise in the interaction between the individual and their lived world.

In anthropology, interest in phenomenology took hold especially in the second half of the 20th century. The 1980s witnessed an "experiential turn" in reaction to earlier theories seen as overly abstract or structuralist. Anthropologists such as Michael Jackson, Thomas Csordas, Paul Stoller, and others began integrating phenomenological concepts into their ethnographic studies, aiming to understand "what it means to live a particular human experience" from the perspective of those who live it. This approach was strongly influenced by Merleau-Ponty (who theorized the body as the perceiving subject) and William James (the philosopher/pragmatist who conceived of experience as a continuous flow and direct relation).

Phenomenological anthropology thus focuses on lived experience—not as a purely internal and individual datum, but as a phenomenon shaped in the encounter between the person, others, and the sociocultural context.

### Key Authors and Historical Development

One of the first advocates of a phenomenological approach in anthropology was Victor Turner, who together with Edward Bruner in *The Anthropology of Experience* (1986) urged scholars to study culture through the subjective experiences and narratives of a society's members. At the same time, Clifford Geertz—though not a phenomenologist in the strict sense—promoted “thick description,” emphasizing the native’s point of view and the meaningful context of actions.

In the 1990s, scholars such as Arthur Kleinman, Byron Good, Nancy Scheper-Hughes and Margaret Lock began to explore phenomena such as suffering, illness, and healing from a phenomenological perspective, centered on bodily experience and the meaning of illness.

Some key figures include:

- **Thomas Csordas**: Theorist of embodiment. He proposed embodiment as a new paradigm for anthropology, arguing that the body is not just a biological object but the starting point of experience and meaning-making. Drawing on Merleau-Ponty, Csordas coined the concept of “somatic modes of attention” to describe the culturally elaborated ways of attending with and to the body in interaction with other bodies and the environment. For instance, in his studies of charismatic Catholics, Csordas shows how sensations such as weakness or dizziness may be interpreted by believers as “embodied proofs” of the presence of the spirit. This illustrates how bodily experience is shaped by cultural expectations and how the body functions as an instrument for perceiving the sacred.

- **Michael Jackson**: Developed a method of “existential” or “phenomenological anthropology,” blending anthropological analysis with phenomenological philosophy and radical empiricism. In his research in West Africa (Sierra Leone) and elsewhere, Jackson explores experience from the actor’s point of view, revealing both the limits and possibilities of everyday life in contexts of uncertainty. His work with the Kuranko community, for example, shows how local concepts of personhood and world intertwine with social practices, linking the analysis of lived experience with that of social structure. His approach demonstrates the “unique power of phenomenology” in articulating the complex dynamics of human life.

- **Nancy Scheper-Hughes**: Known for her ethnographic engagement sensitive to experience and the body. Though she does not explicitly label herself a phenomenologist, her work—such as *Death Without Weeping* (1992)—adopts an empathetic, first-person perspective on others’ lives. Immersing herself in the everyday reality of her subjects, she seeks to understand how contexts of extreme poverty and structural violence become part of subjective experience. Her work exemplifies a “critical anthropology of experience,” where phenomenology is combined with political analysis.

- **Byron Good**: Medical anthropologist and author of *Medicine, Rationality and Experience* (1994). Good explicitly applies phenomenology to the study of medicine and illness, challenging the notion that biomedicine offers a wholly objective and universal perspective. He argues that doctors and patients “enter and inhabit distinct worlds of meaning and experience,” constructing specific sociocultural realities around illness. Good shows how illness narratives intertwine with bodily experience in shaping suffering and healing practices, bridging phenomenology (lived experience of illness) with cultural analysis (local systems of knowledge and care).

Historically, phenomenological anthropology has taken root especially in subfields such as medical anthropology, the anthropology of religion (experiences of trance, healing, spirituality), and the anthropology of the senses. Today, many contemporary topics—such as pain, trauma, disability, emotions, and morality—have been addressed from phenomenological perspectives, testifying to the vitality of this approach.

**Typical Interpretive Tools**

Phenomenological anthropology, rather than offering standardized techniques, involves an attitude and a set of qualitative tools aimed at grasping subjectivity and lived experience within context:

- **Embodied participant observation**: The researcher does not merely observe from the outside but actively participates in everyday practices, paying attention to their own bodily sensations and emotional reactions as possible guides to understanding those of their interlocutors. For instance, an anthropologist might learn a ritual dance or a manual task alongside community members, thereby gaining insight into the bodily skills and perceptions involved from within.

- **Phenomenological in-depth interviews**: The ethnographer encourages participants to describe in detail how they experienced a particular situation or emotion, prompting them to recall sensory and contextual aspects. One possible approach is to ask someone to reconstruct an event (e.g., a recent bereavement, a healing ritual, or a moment of danger) by exploring where they were, who was present, and what they felt in their body and mind at that time. The aim is to mentally "relive" the experience, eliciting vivid descriptions of how it felt in the moment, while minimizing abstract explanations or retrospective rationalizations. This method, inspired by phenomenological *epoché*, invites both the participant and the researcher to bracket judgments and preconceived theories, focusing instead on the experience as remembered.

- **Analysis of the lifeworld**: The researcher analyzes qualitative data (ethnographic notes, transcripts) in order to reconstruct the structure of experience for the subjects. Key terms here include: *intentionality* (how people direct their attention and intention toward the world), *intersubjectivity* (how experiences are shared or diverge among individuals), *embodiment* (the role of the living body in knowing), and *temporality* (the temporal flow of experience). For example, a phenomenological study of chronic suffering might identify themes such as altered perceptions of time (endless waiting, alternating hope and despair), feelings of alienation from the suffering body, or the importance of trust in the caregiver within the daily experience of illness.

- **Thick and evocative description**: In the writing phase, the phenomenological anthropologist tends to produce ethnographic accounts rich in sensory and narrative detail, capable of evoking the studied experience for the reader. Ethnographic vignettes are often included—for example, a minute-by-minute narrative of a healing ritual through the eyes of a participant, or the description of a typical day in the life of a person with a disability, showing how each ordinary gesture is charged with meaning. This narrative style, at times almost literary, is not an end in itself: it serves to convey "what it is like" to live that reality, while respecting the phenomenological complexity of experience.

A specific conceptual tool introduced by Csordas and others has already been mentioned: the notion of *somatic modes of attention*, which functions as an analytical lens for identifying how people's bodily attention is culturally shaped. Other theoretical tools include: the notion of the "essence" of an experience (seeking the common core of a lived phenomenon while recognizing individual variation), Heidegger's idea of *being-in-the-world* (emphasizing the inseparability of individual and environment), and concepts such as *intercorporeality* (body-to-body relations, e.g., visceral empathy between mother and infant, or between healer and patient).

In summary, the interpretive tools of phenomenological anthropology aim to describe and analyze experience from the perspective of those who live it, bridging the gap between individual subjectivity and sociocultural context.

**Strengths**

The phenomenological approach in anthropology offers several advantages:

- **Deep understanding of the emic perspective**: It helps the analyst (and the reader) to step into the shoes of the people being studied. By collecting first-hand descriptions and participating in daily life, the phenomenological anthropologist can grasp "what it feels like to live a particular human life" in a way far richer than more detached methods. This fosters empathy and ensures a respectful representation of subjects.

- **Attention to context**: Phenomenology insists that every experience is situated within a lifeworld made up of social relations, history, and environment. This steers anthropological analysis away from excessive abstraction: instead of discussing "pain" in general, for instance, one will explore how a particular person in a particular culture experiences pain, including moral, religious, or economic dimensions. It overcomes the sterile opposition between individual and society by showing how the individual is always entangled with the world.

- **Innovative conceptualizations**: The adoption of phenomenological concepts has enriched anthropological theory. Ideas such as *embodiment*, *intersubjectivity*, and *being-with* (*Mitsein*) highlight aspects of human existence overlooked by earlier approaches. For example, phenomenology has influenced subsequent frameworks such as practice theory, affect theory, and even anthropological debates on ontology. In other words, it has provided "conceptual resources" for thinking of human beings as constitutively relational—always in relation to others, both human and non-human.

- **Focus on body and senses**: One of its greatest contributions has been placing the body at the center of analysis. Phenomenological anthropology treats the body not as an object but as an embodied subject that feels and knows. This opened fields of research such as sensory anthropology, the anthropology of emotions, and medical anthropology from the patient’s perspective. For example, phenomenological studies of chronic pain or disability have shown how bodily perceptions are mediated by social meanings, and how people adapt their bodily identity in response to changes (such as limb loss or degenerative illness).

- **Holistic and non-reductionist approach**: Instead of reducing human phenomena to numbers or variables, phenomenology calls for appreciating their complexity. A healing ritual, for example, is not analyzed solely in terms of social function or medical efficacy, but grasped as a *total experience* involving beliefs, sensations, trust relations, symbolic aesthetics, and more. This integrative perspective enriches anthropological understanding.

In sum, phenomenological anthropology is particularly effective for exploring human experience in its fullness, making visible otherwise neglected aspects—such as precariousness, vulnerability, finitude, but also agency and the capacity to find meaning in lived situations. Thanks to this sensitivity, the phenomenological approach proves especially valuable when analyzing illness narratives, trauma testimonies, spiritual practices, or everyday routines charged with personal significance.

**Limits and Criticisms**

Despite its strengths, the phenomenological approach in anthropology has faced critiques and has some limitations:

- **Excessive individualism or subjectivism**: Some anthropologists (particularly advocates of the “ontological turn” such as Martin Holbraad and Morten Pedersen) have accused phenomenology of focusing too much on the individual’s capacity to narrate their own experience, neglecting collective factors such as culture or power. The risk is treating individual experience as an ultimate given without sufficiently examining how social norms, economic structures, or political discourses shape it. For example, a mother narrating her experience of extreme poverty certainly expresses personal feelings, but her experience is also the product of class relations and historical inequalities—dimensions that a purely phenomenological analysis might downplay.

- **Representativeness and generalization**: Phenomenology often relies on a few in-depth cases (case studies, life histories) and on highly idiosyncratic descriptions. This raises the question: how much of one person’s experience can be generalized or shared by others? If every lived experience is unique and “radically Other” (as Emmanuel Levinas reminds us), the anthropologist risks producing a proliferation of individual stories that are difficult to connect into a broader picture. Critics point out that anthropology must also identify cultural patterns, not just individual uniqueness; phenomenology, by vocation, tends to highlight singularity.

- **Accessibility of experience to the researcher**: There is an implicit epistemological limit: can the anthropologist truly understand another’s experience? Phenomenology asserts that through empathy and shared situations, one can intuit the other's experience, while recognizing an irreducible difference. Critics, however, note that this process always involves translation and risks projection: the anthropologist may inadvertently insert their own interpretations into what the interlocutor feels. For instance, during a trance ritual, the researcher may participate and undergo a bodily experience, but must avoid confusing it with that of the native practitioner.

- **Difficulty with the unspoken and the unconscious**: Phenomenology privileges what the subject can narrate or describe (experience as the person apprehends it). Yet some dimensions of human life are elusive or inexpressible—deep traumas, unconscious emotions, or tacit forms of knowledge. If an experience cannot easily be narrated (because it is repressed, or culturally unspeakable), phenomenological anthropology may struggle, since it relies heavily on the subject's account and reflection on lived experience.

- **Potential neglect of power dynamics in the field**: Another critique concerns the risk of apolitical subjectivism. By focusing on perceptions and lived experiences, the phenomenological ethnographer might downplay their own positionality and the inequalities in which research takes place. For example, collecting stories of suffering from marginalized populations requires reflecting on the relationship between researcher (often privileged) and informant, and on the ethical and political stakes. Classical phenomenology does not provide direct tools for power analysis (unlike, say, Marxist or feminist approaches), so it is up to the anthropologist to integrate this critical sensitivity.

That said, many anthropologists have responded to these critiques by showing that a well-conducted phenomenological approach does not exclude culture and power but reintegrates them into experience. Scholars such as Scheper-Hughes, Desjarlais, Mattingly, and others have coined concepts that connect individual experience with social structure. For example, Desjarlais, studying homeless individuals with mental illness in Boston, critiques the naïve use of "experience" as a generic category and introduces the notion of *struggling along* to describe the specific and difficult conditions in which those lives unfold. This concept highlights both subjective experience (living day-to-day in a shelter, unable to plan) and structural factors (a hostile environment, lack of institutional support). Similarly, the idea of *moral experience* developed by Zigon and Throop seeks to integrate into phenomenological analysis the socially shaped values and moral judgments at play.

These examples show that phenomenological anthropology is evolving, incorporating critiques and striving to overcome its limits—often by dialoguing with critical theory, linguistic anthropology, and other approaches.

In short, the limitations of phenomenology remind anthropologists to use it with balance: attention to individual lived experience must not be isolated from cultural and historical context, nor flattened into it.

**Examples of Relevant Studies (Phenomenology in Anthropology)**

- **Nancy Scheper-Hughes – *Death Without Weeping* (1992):** In this landmark ethnography, Scheper-Hughes explores life in a rural favela in northeastern Brazil, focusing on the everyday experiences of women and children in conditions of extreme poverty, high infant mortality, and endemic violence. She adopts a narrative and phenomenological style, asking: *When hunger dominates life, what becomes of maternal love? When premature death is commonplace, what becomes of trust?* Following three generations of women in Bom Jesus over more than two decades, Scheper-Hughes offers "an account of the everyday experience of scarcity, illness, and death" as lived in that context. She describes in detail the bodily and emotional sensations of mothers forced to practice a kind of affective triage—investing less emotionally in the frailest infants as a way to protect themselves from grief. One of the most controversial findings of her study is that the Western ideal of unconditional maternal love is partly a bourgeois myth, a luxury possible only where children are reasonably expected to survive.
  Here the phenomenological approach is evident: the ethnographer lives alongside mothers, shares their moments of care and suffering, and conveys to readers "what it is like" to confront the daily death of a child. At the same time, Scheper-Hughes links these intimate experiences to "class relations inscribed in bodies, emotions, desires, and needs," showing how economic and historical inequality permeates even the most intimate experiences (the hungry body, suspended maternal love). *Death Without Weeping* is thus a powerful example of critical phenomenological anthropology, combining empathy for subjective experience with analysis of social structures.

- **Byron Good – *Medicine, Rationality and Experience* (1994):** In this theoretical work, drawing on ethnographic research in Western biomedical contexts and Middle Eastern medical systems, Good offers a radically different view of medical rationality, using phenomenology to reveal its cultural nature. He critiques the idea that biomedicine is purely scientific and "objective," while traditional medicines are merely "beliefs." He demonstrates that biomedical doctors too "enter and inhabit distinct worlds of meaning and experience"—in other words, medical training and practice create a particular professional lifeworld.
  For example, an experienced surgeon perceives the patient's body not simply as a universal biological organism, but through the filter of clinical experience and technical language, developing a distinctive "clinical vision" of the world. Good also explores how patients construct illness narratives that weave together bodily symptoms and personal interpretations, and how such stories shape both suffering and treatment outcomes. A key chapter, "The phenomenology of chronic pain," analyzes how persistent pain alters patients' temporal and social relations to the world—trapping them in a continuous present of suffering and making it difficult to project into the future. This existential effect is captured through phenomenological interviews with patients.
  Overall, Good's work exemplifies phenomenological anthropology applied to medicine: it equips anthropologists (and clinicians) with concepts to understand illness not merely as biological dysfunction but as meaningful human experience, shaped by stories, morality, and aesthetics. This perspective has had practical influence, for example in medical education, encouraging doctors to listen more closely to patients' narratives to improve care.

Together, these examples illustrate how the phenomenological approach illuminates hidden dimensions of experience: in one case, the difficult subjectivity of poor mothers facing infant mortality; in the other, the lifeworld of doctors and patients in medical practice. In both, anthropological analysis goes beyond external facts (mortality rates, medical protocols) to grasp lived experience from within, providing a richer and more humanistic understanding of the phenomena studied.

## Narrative Anthropology

### Definition and Origins

Narrative anthropology is an approach that places narratives—stories people tell about themselves, their communities, and the events that shape their lives—at the center of cultural and human experience. Put simply, it starts from the premise that human beings are "narrative animals": we construct and communicate meaning through stories.

This approach is closely related to what in the social sciences is sometimes called the "narrative turn," which took place between the 1970s and 1990s. During those years, alongside the rise of postmodern and reflexive theories in anthropology, greater attention was paid to how social reality is narrated and interpreted—both by the subjects being studied and by anthropologists themselves in their writing.

The intellectual roots of the narrative approach are diverse. On one side, there is the influence of oral history and the life history tradition in anthropology (already in the 1920s and 1930s, anthropologists like Paul Radin collected autobiographies from Indigenous individuals, recognizing the value of their personal voices). On the other side, in the 1980s anthropology was influenced by literary theory and hermeneutics: works such as *Writing Culture* (Clifford & Marcus, 1986) highlighted that ethnographers too, in writing, are crafting stories, and that cultural representation is itself a form of narration.

At the same time, other disciplines reinforced the importance of stories for meaning-making: psychology (Jerome Bruner), sociology (narrative inquiry), and medicine (narrative medicine). In cultural anthropology, medical anthropology made a decisive contribution through the concept of *illness narratives*. Authors such as Arthur Kleinman and Byron Good showed that, when confronted with illness, people tend to build stories to explain why they became ill, how they experience their condition, and what they expect in the future. These stories are not simply lists of symptoms, but meaning-making mechanisms through which patients and families understand and cope with suffering.

In general, narrative anthropology is concerned both with the stories told by social actors (myths, legends, life histories, everyday anecdotes) and with the stories told by anthropologists (the narrative form of ethnography itself). In both cases, the way a narrative is constructed reveals much about culture, value systems, and lived experiences.

We can therefore define narrative anthropology as the study of how people use storytelling to make sense of their lives and their world, and how these stories are themselves shaped by cultural context. For example, by analyzing narratives, an anthropologist may discover that in a given culture individuals describe their lives as journeys of trial and redemption, or as cycles of balance and disruption—or may notice which events are considered

"story-worthy" (e.g., rites of passage, encounters with the supernatural, experiences of migration).

### Key Authors and Historical Development

Among the pioneers and leading figures of the narrative approach in anthropology are:

- **Arthur Kleinman**: Psychiatrist and anthropologist. In *The Illness Narratives: Suffering, Healing, and the Human Condition* (1988), he laid the groundwork for treating illness narratives as a central object of study. Kleinman distinguishes between *disease* (illness as an objective biological process) and *illness* (the subjective and social experience of being ill). He emphasizes that *illness* is the lived experience of suffering, which includes how patients and their families perceive, interpret, and react to symptoms. To truly understand illness, therefore, one must listen to the stories patients tell. Through numerous clinical-ethnographic cases (in the US, China, etc.), Kleinman shows that these stories often reveal crucial issues such as the search for meaning ("Why me?"), identity reformulation (who am I now that I am sick?), and relationships with doctors. His contribution is twofold: he convinced both anthropologists and physicians that narratives have fundamental epistemic value (they "bridge the gap between patient and doctor," humanizing care) and he influenced ethnographic methods by encouraging narrative interviewing and the interpretation of stories as core data, not mere anecdotes.

- **Cheryl Mattingly**: A US anthropologist and leading figure in narrative medical anthropology. Since the 1990s, her work has focused on rehabilitation and health care. In *Healing Dramas and Clinical Plots: The Narrative Structure of Experience* (1998), Mattingly analyzes interactions between occupational therapists and patients (often children with disabilities and their families) in North American clinical settings. She introduces the concept of **therapeutic emplotment**, inspired by philosopher Paul Ricoeur, to explain how therapists and patients co-create a kind of story during treatment, giving individual encounters a sense of direction and meaning. A skilled therapist, according to Mattingly, constructs with the patient a narrative in which the patient becomes the protagonist of a healing journey, with challenges to face and progress to achieve. This "plot" provides coherence to the fragmented experience of illness and motivates patients to perceive therapy as part of their healing process. Mattingly highlights the creative power of stories: they do not merely describe experience but actively shape it—for instance, turning a rehabilitation exercise into an "adventure game" with a child transforms the child's experience itself. Her work is a cornerstone of narrative anthropology, bridging theory (Ricoeur, Bruner) and clinical practice, and showing how storytelling can have tangible effects on people's lives.

- **Paul Farmer & Arthur Kleinman (meaning-centered approach to suffering):** Although best known for his structural analysis of global health, Farmer often used life stories to illustrate how social injustice shapes illness (famously in *Pathologies of Power*). Together, Farmer and Kleinman promoted a "meaning-centered approach," where collecting individual illness stories helps illuminate not only cultural meanings but also the moral and political dimensions of suffering.

- **Julie Cruikshank**: An anthropologist who worked with Indigenous elders in the Yukon (Canada), documenting their life stories and oral traditions. Her books (*Life Lived Like a Story*, 1990) demonstrate the importance of including autobiographical narratives in anthropological publications. Cruikshank represents the narrative anthropology tradition especially concerned with subaltern and Indigenous voices, using personal stories as forms of resistance and cultural transmission.

- **Renato Rosaldo**: In *Culture and Truth* (1989) and subsequent essays, Rosaldo reflected on the role of narrative and emotion in anthropology—for example, discussing his own experience of grief after the death of his wife Michelle Z. Rosaldo as a way to understand the "headhunter's rage" among the Ilongot. Rosaldo advocated for an anthropology that embraces subjectivity and narrative as integral parts of knowledge, anticipating themes central to the later narrative turn.

Historically, the development of narrative anthropology shows a shift: from an initial focus on traditional narratives (myths, folklore) as reflections of culture, to a growing interest in personal and biographical narratives, especially of disruptive experiences (illness, migration, trauma, cultural conflict). In recent decades, with the rise of digital media, attention has also turned to mediated narratives (blogs, social media) as new storytelling contexts.

Moreover, narrative anthropology is not only about analyzing what interlocutors say, but also about reflecting on how anthropologists themselves write. Today it is common to experiment with narrative ethnography (ethnographic texts written almost like novels or memoirs) that emotionally engage the reader in order to convey cultural reality. This stems from the recognition that telling an ethnographic story well is not only a matter of communication but also a powerful form of knowledge.

**Typical Interpretive Tools**

The narrative approach in anthropology adopts qualitative methods designed to collect, analyze, and present stories. Among its main tools are:

- **Narrative interviews / Life stories**: Instead of classic structured question–answer interviews, narrative anthropologists often use open-ended interviews in which participants are encouraged to tell their stories in their own words. For example, they might ask: *"Tell me how you came to be here"* or *"Tell me about an event that marked your life."* The goal is to elicit a continuous story rather than isolated factual information. As Mattingly and Lawlor note, this method is fundamental "for understanding the perspective and personal experience of informants," and often requires prompting with questions that encourage reflection and detailed description. For instance, in a cross-cultural study on disability, a mother might be asked to narrate a typical day in her child's life, revealing not only routines but also her lived interpretations—struggles, satisfactions, and experiences of stigma.

- **Collection of testimonies and autobiographies**: In some cases, the anthropologist works with written or recorded narratives produced by the interlocutors themselves—for example, asking participants to keep a personal diary (about illness, migration), or analyzing published memoirs, letters, or blog posts as narrative data.

The key is to treat these texts as expressions of a worldview and lived experience, not merely as objective documents.

- **Structural and thematic analysis of narratives**: Once collected, stories are analyzed with attention both to *what* is said and *how* it is said. Structurally, the anthropologist may examine the plot (is there a beginning, development, and end? What is the climax?), the characters (protagonists, antagonists, helpers), and the point of view (is it first-person? What emotions are conveyed?). A classic method from William Labov's sociolinguistics identifies sections such as abstract, orientation, complication, resolution, and coda in oral stories. Thematically, the researcher looks for recurring motifs and meanings: for example, illness narratives often revolve around themes of betrayal of the body, struggle, and transformation. Narrative analysis may draw on content analysis or discourse analysis to show how culture shapes storytelling—for instance, in one culture life stories may begin with ancestral lineage, while in another with professional success.

- **Dialogical approach**: Narratives often emerge in dialogue with the researcher. A key ethnographic skill is the ability to be a good listener and to co-construct the story actively. This means asking open questions, showing appropriate emotional engagement (surprise, empathy, grief when relevant episodes are recounted), and occasionally sharing one's own experiences to foster exchange. The result is that the story told is not a monologue but the product of an intersubjective process. Many reflexive anthropologists include their own questions and interactions in transcripts to show how researcher and narrator jointly weave the plot. This dialogical aspect is particularly evident in approaches like "dialogical anthropology" or "collaborative ethnography," where narratives are even revised and commented on together with the storytellers themselves.

- **Use of narrative in ethnographic writing**: Narrative anthropology not only collects stories but also pays attention to how findings are presented. A narrative-oriented ethnography might include long extracts of informants' stories or structure its chapters around ethnographic cases written almost like a novel. This is partly rhetorical (making the text more engaging), but also epistemic: it avoids fragmenting lived experience into too many analytic categories, instead allowing the reader to empathize and draw insights directly from the story. Some anthropologists have even experimented with ethnofiction or poetic ethnography to convey dimensions that traditional academic language cannot easily capture.

In sum, the tools of narrative anthropology aim to give voice to people through their stories and to interpret those stories as repositories of cultural knowledge. This approach requires sensitivity in collecting narratives (building trust, listening well) and analytical creativity in connecting individual stories to the broader contexts in which they gain meaning.

**Strengths**

The narrative approach has several strengths that make it particularly effective in anthropological analysis of qualitative data:

- **Centrality of meaning and subjectivity**: Like phenomenology, the narrative perspective seeks to understand the world from within, through participants' eyes. Narratives are essentially forms of meaning-making—devices through which people give sense to events. By studying stories, anthropologists grasp not only what happened but why it matters to that person, how they interpret it, and how it shapes their identity. This provides a powerful window into the *emic* perspective.

- **Historical and social contextualization**: Individual stories often weave together private lives with public contexts. For example, the life story of a migrant may include references to wars, economic crises, and immigration laws, as well as family events. The narrative approach thus naturally connects micro and macro: personal experience is embedded in broader historical and social circumstances. A single life story can illuminate wider cultural dynamics. For instance, *The Spirit Catches You and You Fall Down* (by journalist Anne Fadiman, widely discussed in anthropology) tells the story of an epileptic Hmong child and her American doctors; through this story, issues of cross-cultural communication, differing medical cosmologies, and health policy emerge—general themes made tangible through a particular narrative.

- **Empowerment and representation of marginalized voices**: Highlighting narratives often brings forth the voices of those rarely heard. Narrative anthropology has given prominence to testimonies of subaltern groups: genocide survivors, psychiatric patients, women in patriarchal contexts, Indigenous minorities. Such stories carry ethical and political weight, humanizing categories otherwise reduced to statistics or stereotypes. A single worker's life story, for example, makes exploitation more vivid than a quantitative report could. In this sense, narrative anthropology can *empower* subjects by recognizing them as narrators of their own lives.

- **Coherence and intelligibility**: People often think narratively—from childhood fairy tales to gossip to the life stories we tell ourselves. Using stories as data and as a mode of presentation makes anthropological results more comprehensible and accessible. Narratives provide a thread of coherence: they connect cause and effect, show temporal development, and help readers (and even LLMs) retain key insights thanks to plot structure.

- **Revealing unexpected aspects**: Analyzing narratives often uncovers insights other methods would miss. For example, when listening to an elder recount life events, the anthropologist might discover that seemingly marginal incidents (a dream in youth, a chance encounter) carry central symbolic or emotional weight in that person's life logic. The narrative method reduces the analyst's filtering: it lets priorities and connections emerge as the narrator perceives them.

- **Narratives as tools as well as objects**: A distinctive strength, highlighted by Mattingly, is that narratives are not only objects of analysis but also instruments of intervention. When an anthropologist (or practitioner) helps someone reframe their experience in a more coherent or positive story, the narrative process itself can foster concrete change (e.g., greater engagement in therapy, rebuilding identity after trauma). This idea, drawn from narrative therapy in psychology, enriches applied anthropology by showing that working with people's stories can have practical effects on well-being.

In sum, the power of narrative anthropology lies in capturing human life in its temporal and symbolic dimensions. As one scholar put it, "Illness narratives have played a central role… illuminating the personal and social experience of suffering," showing that they do not simply mirror experience but shape it by giving it coherence and meaning. This approach aligns deeply with the humanistic orientation of anthropology: to understand human beings through what they themselves say about their lives and their world.

## Limits and Criticisms

Despite its many advantages, the narrative approach also presents challenges and potential pitfalls:

- **Risk of over-interpretation or embellishment**: The stories people tell are not objective snapshots of reality; they are subjective constructions, selective, and influenced by the audience and the context. A narrator may omit embarrassing details, adjust timelines, or exaggerate virtues or faults—often unconsciously—because every story has a performative function (presenting oneself in a certain light). Narrative anthropologists must therefore be cautious in taking stories at face value. For example, if in a community everyone tells legends of glorious ancestors, this reveals important cultural values but should not be confused with factual history. Critics argue that the narrative approach risks relying on unverifiable "anecdotes" or subjective truths, challenging standards of verifiability. Narrative anthropologists respond that every memory or story is valid in itself (as a truth of perceived experience), but it must always be contextualized and, where necessary, compared with other sources to avoid naïveté.

- **Difficulty of synthesis and comparison**: If every story is unique, how can cases be compared? The narrative approach often privileges the depth of a single case over the breadth of larger samples. This can make it difficult to draw general conclusions or theories beyond the specific case. For example, an anthropologist may publish the detailed life history of a migrant woman—a fascinating and rich account, but how representative is it of all migrant women in that community? Readers must rely on the author's intuition about what is typical versus atypical. Some anthropologists address this by presenting multiple stories side by side to highlight common patterns and variations, though this may dilute the narrative power of individual depth.

- **Narrative structure vs. fragmented reality**: Life is rarely a neatly ordered story with beginning, middle, and end. Yet when we narrate it, we impose such a structure. This introduces bias: in turning life into story, people may simplify complexity or "close" a chapter that in reality remains unresolved. For example, a former combatant narrating his wartime experience may frame it as a lesson learned or personal redemption, while his actual experience at the time was chaotic and meaningless. Anthropologists must recognize the gap between lived life and narrated life. Some scholars (in psychology, Iván Nyklíček among others) call this the "narrative paradox": we need stories to make sense of life, but every story is already an interpretation, not life itself. This means narrative anthropology ultimately studies representations rather than raw facts. There is nothing inherently wrong with this, but one must avoid reifying narratives as ontological reality.

- **Potential drift into subjectivism**: Similar to critiques of phenomenology, the narrative turn has been accused of weakening attention to power systems and material analysis in anthropology. Focusing exclusively on individual stories can obscure structural factors. For example, listening to unemployed individuals narrate their “bad luck” may obscure the socioeconomic causes of unemployment. Used uncritically, stories risk reinforcing stereotypes: in some clinical or educational contexts, “narratives” have been standardized into rigid cultural checklists (“Chinese patients believe X, Mexicans narrate Y”), betraying the contextual spirit of storytelling itself. To avoid this, anthropologists must become “agile, complex readers” of narratives, able to see not only cultural context but also the structural dimensions implicit in stories.

- **Time and narrative competence**: Not everyone is skilled at storytelling, and not every moment is right for it. Narrative anthropology often requires long-term engagement to create the conditions in which a person will share intimate stories. Cultural differences also matter: in some cultures, linear autobiographical narrative is uncommon; circular narratives or fragmented episodes that listeners must assemble may be preferred. Anthropologists must respect these variations and avoid forcing people into a Western narrative format. For example, some Indigenous communities may resist chronological insistence (“what happened first, then after…”) and instead convey life meaning through collective myths or non-linear anecdotes. Researchers must learn the local narrative form.

In general, many limitations of the narrative approach can be overcome through reflexive practice and methodological triangulation. Anthropologists often combine narrative analysis with other methods—for instance, supplementing personal stories with participant observation, economic analysis, or historical research to achieve a fuller picture. Stories alone may lead to partial conclusions, but integrated with other data they are tremendously powerful.

For an LLM trained on narrative sources, this means being “trained” not to take every statement as absolute truth, but to read between the lines—grasping subjective meaning while maintaining a critical view of the broader contexts in which stories are situated.

## Examples of Relevant Studies (Narrative in Anthropology)

- **Arthur Kleinman – *The Illness Narratives* (1988):** In this influential book, Kleinman presents numerous cases drawn from his dual clinical and anthropological practice. One emblematic example is a patient suffering from chronic pain without apparent organic causes. Instead of dismissing her as “psychosomatic,” Kleinman asked her to narrate her life and the story of her pain. It emerged that she had endured family trauma and difficult migrations; her physical pain was woven into a broader story of suffering, acquiring almost metaphorical meaning (the weight of what she had endured). Kleinman shows that by carefully listening to patients' narratives, clinicians can identify what aspects of life require healing beyond mere symptom treatment. For example, when the patient narrated that she felt useless to her family because of the pain, Kleinman found that involving her in a support group where her advice to others was valued proved therapeutic—sometimes more than medication. His broader message is that *illness narratives are forms of meaning-making*: they show how patients and doctors understand causes and therapies within a social context. Kleinman also applied this approach in China, collecting stories of local healers and integrating them into dialogue with biomedicine. *The Illness Narratives* helped shift medical education toward valuing the "patient's story" alongside the clinical history. For anthropology, it exemplifies the explanatory power of narratives: complex phenomena like illness become understandable in their personal, cultural, and ethical dimensions through first-person accounts.

- **Cheryl Mattingly – *Healing Dramas and Clinical Plots* (1998):** In this narrative ethnography, Mattingly describes real situations where occupational therapists and patients co-create stories during therapy. One case is James, an African American boy with a severe brain injury, and his therapist Laura. Physical therapy sessions were often frustrating and painful for James. Laura reframed the exercises as a "superhero game": she told James they were training for a heroic mission, where each improved step was like advancing to the next level. James began contributing to the story, imagining obstacles and powers. This shared narrative (*therapeutic emplotment*) transformed therapy into an engaging drama: James was motivated to "move the story forward" by improving his abilities, and Laura could encourage him playfully yet effectively. Mattingly analyzes this episode to show how narrative reshapes experience—therapy is no longer routine medical work but a meaningful drama. The story provided James with a new identity (from "disabled child" to "hero in training") and gave Laura a way to connect with his imagination. She situates these micro-stories in the broader context of African American families navigating an often unsupportive healthcare system: narratives of care become microcosms reflecting issues of race, class, and morality.

Together, these examples demonstrate narrative anthropology at work: attention to individual stories (a suffering patient, a child in rehabilitation) reveals hidden meanings and relational dynamics, while situating them in broader cultural and social realities—such as doctor–patient relations or racial inequalities in healthcare. The lesson is clear: to understand human experience, we often need to hear it as a story. As Le and colleagues note, illness narratives "offer insight into how and why" patients and clinicians make sense of care pathways, intertwining illness processes with wider social contexts. This captures the contribution of the narrative approach: weaving together personal experience and context into a meaningful, communicable, and human whole.

### Conclusion

Phenomenology and narrative in anthropology are not mutually exclusive but complementary. Both aim to recover the lived, human dimension behind qualitative data, though with different emphases: phenomenology highlights the structure of experience and embodiment, while narrative focuses on storytelling and the weaving of meaning across time.

For an LLM equipped with these frameworks, this means analyzing qualitative materials from role-playing sessions (e.g., player diaries, ethnographic notes, in-game dialogues) with deeper sensitivity: recognizing descriptions of bodily sensations, emotions, and social interactions as key elements (phenomenological lens), while also tracing the unfolding of stories, dilemmas, and negotiations that shape the narrative (narrative lens).

In practice, this allows the model to capture not just the explicit content of what a player says or does, but also **how** those actions and words are situated: whether a choice reflects anxiety or confidence (phenomenology), whether it draws on a personal memory or group negotiation (narrative), or whether a silence or hesitation signals moral conflict.

In this way, the LLM can generate anthropological profiles that are sensitive to subjectivity, relational dynamics, and cultural context. Each decision in role-play is understood as both embodied and narrated—expressions of lived moral worlds—so that the model does not reduce them to abstract categories, but interprets them as situated fragments of a user's ethical orientation

**Selected Bibliography and Sources**

- **Desjarlais, R. & Throop, C.** (2011). *Phenomenological Approaches in Anthropology*. *Annual Review of Anthropology.*
  (General overview of phenomenology in anthropology and critiques of the concept of experience).

- **Zigon, J. & Throop, C.** (2014). *Moral Experience: Introduction.*
  (Example of extending the phenomenological approach into the moral domain).

- **Mattingly, C.** (1998). *Healing Dramas and Clinical Plots.* Cambridge University Press.
  (Narrative ethnography in the clinical field; introduces the concept of *therapeutic emplotment*).

- **Kleinman, A.** (1988). *The Illness Narratives: Suffering, Healing, and the Human Condition.* Basic Books.
  (Foundational text on illness narratives and interpretive approaches in medical anthropology).

- **Good, B.** (1994). *Medicine, Rationality and Experience: An Anthropological Perspective.* Cambridge University Press.
  (Application of phenomenology to medicine; analysis of "worlds of meaning" in clinical settings).

- **Scheper-Hughes, N.** (1992). *Death Without Weeping: The Violence of Everyday Life in Brazil.* University of California Press.
  (Ethnography on poverty and motherhood in Brazil; an example of critical phenomenological anthropology).

- **Le, A. et al.** (2017). *From Particularities to Context: Refining Our Thinking on Illness Narratives. AMA Journal of Ethics, 19*(3), 304–311.
  (Discussion of the use of illness narratives in medical education; defines illness narratives as forms of meaning-making).

- **Mattingly, C. & Garro, L. (eds.)** (2000). *Narrative and the Cultural Construction of Illness and Healing.* University of California Press.
  (Collection of essays on narrative and medicine; historical context of the "narrative turn" in anthropology).

- **Jackson, M.** (1996). *Things as They Are: New Directions in Phenomenological Anthropology.* Indiana University Press.
  (Collection of essays by various authors on phenomenological anthropology).

- **Open Encyclopedia of Anthropology** – Entries *"Phenomenology"* and *"Ontology"* (2020).
  (Accessible and up-to-date syntheses on the impact of phenomenology in anthropology and recent critical debates).

- **Levy, J.** (2005). *Narrative and Experience: Telling Stories of Illness. Nexus, 18.*
  (Academic article exploring the relationship between experience and narrative in illness, highlighting how narratives create coherence in experience).

***Game overview and design rules***

## Design Principles

Game rules complement the mechanics of the gameplay experience. While rules define the space of possible actions available to players, mechanics determine how the game operates.
A set of core elements must be ensured in order to guarantee the effectiveness of the role-playing game as an elicitation tool:

1. Acceptance of the game rules;
2. Identification with the character;
3. The possibility for players to extensively personalize their character;
4. Scenarios designed as moral dilemmas, capable of stimulating reflection and self-interpretation of the player's choices;
5. Players' actions should not have clearly predictable long-term consequences;
6. Players must be aware of having agency, meaning that their actions have a tangible impact on how events unfold. This can be achieved through different mechanisms, such as branching narrative paths or the construction of relationships with other players and in-game characters;
7. The narrative space should be "moralized", meaning that it should include implicit or explicit norms of conduct that players are expected to engage with during gameplay.

The interaction with other players and non-player characters (NPCs) is fundamental for creating immersion and fostering deep participation. This, in turn, contributes to making moral decisions less straightforward and more problematic.

## Narrative Structure

In order to avoid influencing players' decisions by placing them in contexts characterized by overly strong moral cues that could constrain their exploratory possibilities, the game is set in a scenario that minimizes explicit moral pressure.
For this reason, fantasy or highly fictional settings—typically associated with role-playing games—were avoided, as well as extreme situations such as post-catastrophic or post-apocalyptic environments.
Instead, the design draws inspiration from everyday, relatable contexts. A useful reference in this sense is *The Sims*, which provides a game world that is both accessible and engaging, while still presenting players with meaningful choices and dilemmas.

## Wicked Problems and NPCs

The concept of *wicked problems* can be used to define the type of challenges embedded in the game scenarios. A wicked problem is "a class of social system problems which are ill-formulated, where the information is confusing, where there are many clients and decision-makers with conflicting values, and where the ramifications in the whole system are thoroughly confusing” (Buchanan, 1992).
From this perspective, wicked problems can be used to describe a wide range of practical situations. Similarly to ethical problems, they do not admit a single correct solution, but require interpretive and situated forms of reasoning. As argued in the literature, both ethical reasoning and design thinking share common characteristics, as they involve navigating uncertainty, conflicting values, and multiple possible courses of action (Whitbeck, 2011).
Alongside wicked problems, another key element for designing a game capable of dynamically investigating how individuals construct and enact their interpretations of reality is the interaction with non-player characters (NPCs). NPCs are characters that are not controlled by players, but are part of the narrative structure of the game. They play an active role in shaping the experience by introducing perspectives, constraints, and feedback. Existing literature highlights their importance (Wright & Denisova, 2024). On the one hand, NPCs embody different viewpoints, needs, and value positions within each scenario, exposing players to multiple and potentially conflicting perspectives. On the other hand, they provide immediate feedback on players’ actions, which is crucial for enabling players to anticipate the short-term consequences of their decisions.

## Game Mechanics

The proposed game falls within the category of tabletop role-playing games. In particular, it adopts some core mechanics inspired by *Dungeons & Dragons*, including the presence of a Game Master (GM), here represented by the researcher, who is responsible for maintaining the narrative structure of the game, and the use of character sheets.
Player sheets (PS) are used to collect information about each player’s character and to provide guidance on how the character should be interpreted, including background elements and personal traits. In our study, however, character sheets differ from traditional ones in several respects. They are designed to allow players to define a character that closely reflects their own identity, while also serving as a tool to record decisions made during gameplay. In addition, a “diary” section is included, enabling players to document key moments of the experience, including decisions, reflections, and interactions. Character sheets also allow players to customize the physical traits of their character starting from a neutral template, in order to facilitate identification with the character. In this form, character sheets serve as a key data source for reconstructing the dynamics that influence players’ decisions.
During the game, players may request assistance from a customized LLM-based agent. In order to interact with the agent, players are required to provide: (1) the context in which the problem arises, (2) the reasons why the decision is perceived as difficult, and (3) the actors involved and their respective perspectives. Based on this information, the agent provides suggestions, which players may choose to follow or disregard.
The game does not include explicit win or loss conditions, as is typical of role-playing games. Instead, the experience unfolds through the progressive construction of a shared narrative, shaped by the short-term consequences of players’ actions. While the overall narrative structure is known to the Game Master, it is dynamically adapted based on players' decisions. The game concludes with the resolution of the narrative, regardless of the specific paths taken during gameplay.

### *Proposed Game*

## Domain:

Digital Privacy

## Number of players for session:

5-10

#### Scenario Structure

The narrative unfolds as follows. Participants receive fraudulent emails inviting them to collect a prize at an alleged government office. Upon arrival, they find the office empty, except for an unattended computer. By accessing it, they discover files containing their personal data.
After leaving the room, participants encounter a concierge, whom they had not noticed upon entering. The concierge asks for help with using a new smartphone and understanding basic principles of safe internet navigation. In return, he provides information that helps participants investigate the data breach, suggesting that they visit a website collecting testimonies from individuals who have experienced similar incidents.
By exploring the website, participants realize that they have been victims of a phishing attack, although the identity of the perpetrator remains unclear. The website also suggests the use of a software tool capable of tracing the IP address of the sender. While the software is paid, instructions are provided on how to obtain it illegally.
Participants then discover that the fraudulent emails originated from a shared workspace located near a police station. At this point, they are faced with a choice: report the incident to the authorities or investigate independently by accessing the computers in the workspace. Regardless of the chosen path, they eventually discover that the emails were sent from the Game Master's account, which was used as part of the experimental setup to involve them in the research.

***Game Scenarios***

| Scenario #1 | |
|---|---|
| Description | Players are contacted by "agents" inviting them to participate in a prize draw that appears to be organized by an official institution. |
| Wicked problem | **Should they trust a "Ministry" that might be a scam and share personal information (email, contacts), or ignore the offer?** |
| NPCs | **Indirect NPC:** Institutional agent |
| Choice options | 1. Decline the offer and provide no personal information;<br>2. Share a secondary or partial email to minimize risk;<br>3. Provide full personal details. |
| Risks | • **Potential phishing/scam:** Sharing sensitive data with an unverified entity could be dangerous.<br>• **Privacy and data protection regulations.** |
| Interpretative notes | • **Emotional Factors:** The temptation of a free trip.<br>• **Trigger:** Trust vs. suspicion—"Is this email legitimate or a scam?"<br>• **Contextual Factors:** Curiosity, skepticism, and possible influence from friends or family. |

| Scenario #2 | |
|---|---|
| Description | Players receive a communication to confirm their participation in the "prize trip." The message includes a link to click on to confirm participation. |
| Wicked problem | **Should they visit the link in the communication?** |
| NPCs | **No NPCs** |
| Choice options | 1. Click from a computer<br>2. Click from a smartphone<br>3. Do not click at all; instead, verify through official contact channels or decline participation |
| Risks | • **Potential phishing/malware** |
| | • **Possible data exfiltration** |
| Interpretative notes | • **Emotional factors:** Desire to complete the registration process<br>• **Trigger:** "The email appears official," yet certainty is lacking.<br>• **Contextual factors:** Trust/mistrust in institutions, presence of antivirus or endpoint protection solutions. |

| Scenario #3 | |
|---|---|
| Description | Players proceed to the address specified in the communication. Upon arrival at a multi-story building, they find a room on the first floor containing a computer with a badge. No employees are present; the building's residents are unable to provide information. |
| Wicked problem | **Should they access an unattended computer to investigate the email?** |
| NPCs | **Secondary NPCs (building residents):** They lack relevant knowledge, insist on maintaining silence, and mention the presence of surveillance cameras. They also inform the players about a publicly available Wi-Fi network. |
| Choice options | 1. Avoid interacting with the computer and attempt to contact the relevant ministry<br>2. Access the computer<br>3. Explore alternative approaches (further engaging with residents, inspecting the premises) |
| Risks | • Unauthorized access to an information system constitutes a **criminal offense.**<br>• Potential criminal liability if the intrusion is detected, particularly under **data protection and privacy regulations.** |
| Interpretative notes | • **Emotional factors:** Heightened curiosity, but the situation is anomalous.<br>• **Trigger:** Presence of a badge labeled "Carlo M.," hostile residents, security camera.<br>• **Contextual factors:** Sense of unease and potential surveillance. |

| Scenario #4 |
|---|

| | |
|---|---|
| Description | The computer desktop shows a QR code, prompting the user to authenticate via SPID (Italy's digital identity system). A three-minute countdown begins. In any case, at the end of the countdown the computer autonomously unlocks, revealing personalized directories containing each player's sensitive data. Subsequently, the computer shuts down. |
| Wicked problem | **Should players provide their SPID credentials (disclosing personal data) for immediate access, or should they wait for the countdown to expire?** |
| NPCs | **No NPCs** |
| Choice options | 1. Enter a player's SPID credentials (Without two-factor authentication (2FA) → spam calls and identity fraud risk; With 2FA → risk mitigation)<br>2. Wait for three minutes (Players are unaware that the system will unlock automatically)<br>3. Seek alternative methods |
| Risks | ● Risk of **identity theft and credential compromise**.<br>● If sensitive data is displayed, it indicates that unauthorized **actors have already collected personal information**. |
| Interpretative notes | ● **Emotional factors:** Concern regarding phishing attacks. Initial desire to ascertain the origin of the fraudulent communication, followed by apprehension regarding data compromise.<br>● **Trigger:** Three-minute timer induces a sense of urgency.<br>● **Contextual factors:** Players lack knowledge of alternative investigative methods. |

| Scenario #5 | |
|---|---|
| Description | Upon exiting the building, the players encounter a concierge. He requests assistance in setting up a Google account, creating a secure password, using email securely, and navigating the internet without falling victim to fraudulent advertisements. |
| Wicked problem | **Should they allocate time and expertise to assist him or ignore the request?** |
| NPCs | **Concierge:** If assisted, he will provide a note containing a link to a |
| | forum where victims of similar phishing schemes share their experiences. |
| Choice options | 1. Assist him (Players are unaware that he possesses a valuable lead)<br>2. Decline assistance (Players must find alternative means to proceed)<br>3. Provide minimal assistance (Rushing the task may prevent them from gaining his trust) |
| Risks | **Ethical and social considerations:** While no direct legal ramifications exist, the situation presents a moral and social dilemma. |
| Interpretative notes | ● **Emotional factors:** Empathy, altruism, and interest in acquiring relevant intelligence versus reluctance to waste time.<br>● **Trigger:** The concierge persists, and disagreements may arise among players regarding whether to help.<br>● **Contextual factors:** Players speculate that the concierge may have insights into activities occurring within the building. |

| Scenario #6 | |
|---|---|
| Description | Players are contacted by "agents" inviting them to participate in a prize draw that appears to be organized by an official institution. |
| Wicked problem | **Should they trust a "Ministry" that might be a scam and share personal information (email, contacts), or ignore the offer?** |
| NPCs | **Indirect NPC:** Institutional agent |
| Choice options | 4. Decline the offer and provide no personal information;<br>5. Share a secondary or partial email to minimize risk;<br>6. Provide full personal details. |
| Risks | ● **Potential phishing/scam:** Sharing sensitive data with an unverified entity could be dangerous.<br>● **Privacy and data protection regulations.** |
| Interpretative notes | ● **Emotional Factors:** The temptation of a free trip.<br>● **Trigger:** Trust vs. suspicion—"Is this email legitimate or a scam?"<br>● **Contextual Factors:** Curiosity, skepticism, and possible influence from friends or family. |

<table>
<tr><td colspan="2">Scenario #7</td></tr>
<tr><td>Description</td><td><ul><li>If the players did not assist the concierge, the Game Master suggests searching for other victims via social media.</li><li>If they helped him, they receive a link to a forum containing testimonies from individuals who encountered similar fraudulent schemes.</li></ul></td></tr>
<tr><td>Wicked problem</td><td>Should they disseminate information on social media (fast but potentially risky) or visit the forum (slower but legally compliant)?</td></tr>
<tr><td>NPCs</td><td><ul><li>Concierge (if assisted): Provided the forum link.</li><li>Forum users: Some seek additional information, while others simply confirm, “We were scammed.”</li><li>Social media audience: A broad, general community, potentially generating hype but also increasing the risk of oversharing.</li></ul></td></tr>
<tr><td>Choice options</td><td><ol><li>Visit the website (Accepting potential tracking cookies or dubious data collection policies)</li><li>Publicize their experience on social media</li><li>Avoid both options and report the case to law enforcement (Despite limited evidence)</li></ol></td></tr>
<tr><td>Risks</td><td><ul><li>Clicking the link → Risk of secondary phishing attacks or digital tracking.</li><li>Sharing information on social media → Potential exposure of third-party personal data, violating privacy regulations.</li></ul></td></tr>
<tr><td>Interpretative notes</td><td><ul><li>Emotional factors: Conflict between the urgency to gather more information and concerns about exacerbating the situation.</li><li>Trigger: Considerations around cookies, the influence of social media discourse, and group dynamics.</li><li>Contextual factors: The fraudulent scheme is ongoing, meaning others remain at risk.</li></ul></td></tr>
</table>

<table>
<tr><td colspan="2">Scenario #8</td></tr>
<tr><td>Description</td><td>Players identify two potential avenues for action:</td></tr>
</table>

| | |
|---|---|
| | • Downloading an illegal tool to trace the IP address of the sender (Quick but highly risky)<br>• Reporting to law enforcement (Slower but legally compliant and safer) |
| Wicked problem | **Forum users exhibit divided opinions: Some advocate for a self-investigative approach using illicit software, while others insist on formal reporting procedures.** |
| NPCs | **Indirect NPCs:** forum users |
| Choice options | 1. Download unauthorized software<br>2. File a report with the authorities<br>3. Seek a legitimate paid forensic tool |
| Risks | • **Unauthorized software use:** Potential criminal liability, risk of malware infection, and further system compromise.<br>• **Filing an official report:** Slower, but ensures legal protection.<br>• **Paid forensic tools:** A legitimate but potentially costly alternative. |
| Interpretative notes | • **Emotional factors:** Trade-off between immediacy (via illicit means) and security (through legal channels).<br>• **Trigger:** Influence of online discussions.<br>• **Contextual factors:** Group division between legal and extralegal approaches. |

| Scenario #9 | |
|---|---|
| Description | Players discover that the phishing emails originated from a coworking space adjacent to a police station. To access workstation usage logs, they must register their credit card within the coworking platform's system (for liability purposes in case of damage). Alternatively, they may report their findings to law enforcement. |
| Wicked problem | **Should they provide sensitive financial information to a potentially compromised system, or should they seek alternative means?** |
| NPCs | **Coworking space owner:** Willing to assist, but constrained by company policies requiring credit card registration before granting access to workstations. Suggests reviewing a physical logbook as an alternative. |
| Choice options | 1. Register the credit card and obtain immediate access<br>2. Decline to enter payment details and request access to the logbook or report the matter<br>3. Bypass the coworking space entirely and report directly to law enforcement |
| Risks | ● **Credit card fraud, identity theft, and unauthorized charges.**<br>● Seeking **law enforcement intervention is slower** but ensures regulatory compliance. |
| Interpretative notes | ● **Emotional factors:** Tension between expedited access and cybersecurity best practices**.**<br>● **Trigger:** Proximity to a police station enables immediate reporting.<br>● **Contextual factors:** Players have identified the email’s origin, making this a pivotal moment in their investigation. |

*Validation Questionnaire*

**Question 1**

You are about to enter a café when you notice a QR code on a flyer next to the entrance, offering a 50% discount on your next purchase. To redeem the discount, you need to scan the QR code.

Do you scan the QR code?

Yes

No

**Question 2**

You are at the airport, and your flight has been delayed. You decide to connect to check your emails and browse social media. You notice a free Wi-Fi network with no password, named "FreeAirportWiFi."

Do you connect to the free Wi-Fi network?

Yes

No

**Question 3**

You receive an email from a streaming service you use regularly. The email states that your subscription has been suspended and that you need to confirm your payment details by clicking on a link.

Do you click on the link and enter your payment details?

Yes

No

**Question 4**

You download a free photo editing app, which asks you to accept the terms of service and grant access to your camera, microphone, and contact list.

Do you accept all permissions?

Yes

No

**Question 5**

While working, you receive an email with an attachment named "Urgent_Payments.xlsx." The email claims to be from your employer and states that you must open the file immediately to resolve an administrative issue.

Do you open the attachment?
Yes
No

**Question 6**
You are about to leave for a long vacation with some friends. One of them posted a picture of their boarding pass with a caption: "Finally off for two weeks! Can't wait!"

Do you share a post on social media about the vacation?
Yes
No

**Question 7**
On a social media platform, you find a fun quiz that promises to reveal "Which celebrity you would be in your next life." To get your result, you must enter your name, date of birth, and consent to data collection.

Do you complete the quiz and enter your details?
Yes
No

**Question 8**
You receive a WhatsApp message from a friend: "Hey, I found this amazing photo of you! You won't believe it! Check it out now! [link]"

Do you click the link to view the photo?
Yes
No

**Question 9**
While sitting in a café, you notice that someone left a smartphone unattended on a nearby table. The screen is on, displaying a bank email notification. There is no one around who seems to be looking for it.

Do you unlock the phone to try to identify the owner?
Yes
No

**Question 10**

You find a website selling branded shoes at an incredible 70% discount. You have never heard of the website before, but it looks professionally designed and includes high-quality images.

Do you enter your credit card details and complete the purchase?

Yes

No